\documentclass[fleqn,usenatbib]{mnras}

\usepackage{newtxtext,newtxmath}
\usepackage[T1]{fontenc}
\usepackage{ae,aecompl}
\usepackage{graphicx}	% Including figure files
\usepackage{amsmath}	% Advanced maths commands
\usepackage{amssymb}	% Extra maths symbols
\usepackage{multicol}        % Multi-column entries in tables
\usepackage{bm}		% Bold maths symbols, including upright Greek
\usepackage{pdflscape}	% Landscape pages
\usepackage{soul}
\usepackage{enumitem}
\usepackage{pdflscape}	% Landscape pages
\usepackage[dvipsnames]{xcolor}
\newcommand{\Myr}{\mathrm{Myr}}

\newcommand{\kpc}{\mathrm{kpc}}
\newcommand{\pc}{\mathrm{pc}}
\newcommand{\cmsquare}{\mathrm{cm}^{-2}}
\newcommand{\cmcube}{\mathrm{cm}^{-3}}
\newcommand{\protonmass}{m_\mathrm{p}}
\newcommand{\mppercmcube}{\protonmass \, \cmcube}

\newcommand{\Msol}{\textup{M}_\mathrm{\sun}}
\newcommand{\Zsol}{\textup{Z}_\mathrm{\sun}}
\newcommand{\xMsol}[2]{\ensuremath{{#1}\times 10^{#2} \,\Msol}}
\newcommand{\xScientific}[2]{\ensuremath{{#1} \times 10^{#2}}}
\newcommand{\Msolyr}{\textup{M}_\mathrm{\sun} \, \text{yr}^{-1}}
\newcommand{\kms}{\text{km} \, \text{s}^{-1}}

\newcommand{\K}{\mathrm{K}}

\newcommand{\rvir}{r_{200}}

\newcommand{\SFR}{\mathrm{SFR}}
\newcommand{\Mstar}{M_{\star}}

\newcommand{\lcool}{l_{\text{cool}}}
\newcommand{\vz}{v_{z}}
\newcommand{\vcirc}{v_{\text{circ}}}
\newcommand{\massloading}{\eta_{M}}
\newcommand{\energyloading}{\eta_{E}}
\newcommand{\metalloading}{\eta_{Z}}

\newcommand{\nh}{n_{\mathrm{H}}}

\newcommand{\hi}{H\textsc{i}}
\newcommand{\hmol}{\mathrm{H}_\mathrm{2}}

\newcommand{\halpha}{H$\alpha$}

\newcommand{\oi}{O~\textsc{i}}

\newcommand{\oiii}{O~\textsc{iii}}

\newcommand{\ovi}{O~\textsc{vi}}
\newcommand{\ovii}{O~\textsc{vii}}

\newcommand{\ci}{C~\textsc{i}}
\newcommand{\cii}{C~\textsc{ii}}

\newcommand{\civ}{C~\textsc{iv}}
\newcommand{\sI}{S~\textsc{i}}

\newcommand{\siI}{Si~\textsc{i}}
\newcommand{\siII}{Si~\textsc{ii}}

\newcommand{\mgII}{Mg~\textsc{ii}}
\newcommand{\feI}{Fe~\textsc{i}}
\newcommand{\feII}{Fe~\textsc{ii}}
\newcommand{\neII}{Ne~\textsc{ii}}

\title[Resolving galactic outflows]
{Boosting galactic outflows with enhanced resolution}

\author[M. P. Rey et al.]
{Martin P. Rey,\thanks{E-mail: \href{martin.rey@physics.ox.ac.uk}{martin.rey@physics.ox.ac.uk}} Harley B. Katz, Alex J. Cameron, Julien Devriendt, Adrianne Slyz 
\vspace{0.8mm}
\\
Sub-department of Astrophysics, University of Oxford, DWB, Keble Road, Oxford OX1 3RH, UK \\ 
}

\date{Submitted to MNRAS}

\pubyear{2022}

\begin{document}
\label{firstpage}
\pagerange{\pageref{firstpage}--\pageref{lastpage}}
\maketitle

\begin{abstract}
We study how better resolving the cooling length of galactic outflows affect their energetics. We perform radiative-hydrodynamical galaxy formation simulations of an isolated dwarf galaxy ($\Mstar=10^{8}\, \Msol$) with the \textsc{ramses-rtz} code, accounting for non-equilibrium cooling and chemistry coupled to radiative transfer. Our simulations reach a spatial resolution of $18 \, \pc$ in the interstellar medium (ISM) using a traditional quasi-Lagrangian scheme. We further implement a new adaptive mesh refinement (AMR) strategy to resolve the local gas cooling length, allowing us to gradually increase the resolution in the stellar-feedback-powered outflows, from $\geq 200 \, \pc$ to $18 \, \pc$. The propagation of outflows into the inner circumgalactic medium (CGM) is significantly modified by this additional resolution, but the ISM, star formation and feedback remain by and large the same. With increasing resolution in the diffuse gas, the hot outflowing phase ($T > \xScientific{8}{4} \, \K$) systematically reaches overall higher temperatures and stays hotter for longer as it propagates outwards. This leads to two-fold increases in the time-averaged mass and metal outflow loading factors away from the galaxy ($r=5\, \kpc$), a five-fold increase in the average energy loading factor, and a $\approx$50 per cent increase in the number of sightlines with $N_{\text{\ovi}} \geq 10^{13}\, \cmsquare$. Such a significant boost to the energetics of outflows without new feedback mechanisms or channels strongly motivates future studies quantifying the efficiency with which better-resolved multiphase outflows regulate galactic star formation in a cosmological context. 
\end{abstract}

\begin{keywords} 
  galaxies:evolution -- methods: numerical -- hydrodynamics
\end{keywords}

%%%%%%%%%%%%%%%%%%%%%%%%%%%%%%%%%%%%%%%%%%%%%%%%%%

%%%%%%%%%%%%%%%%% BODY OF PAPER %%%%%%%%%%%%%%%%%%

\section{Introduction} \label{sec:intro}

Multiphase galactic outflows are ubiquitous across the galaxy population, with their range of molecular, cold, ionized and hot phases detected across the electromagnetic spectrum (see e.g. \citealt{Veilleux2020, Laha2021} for reviews). By removing mass from the dense ISM and heating the gas surrounding galaxies in the CGM, outflows provide a fundamental mechanism to regulate the star formation of galaxies across cosmic time. But a detailed understanding of how galactic outflows launch, how they propagate through the ISM and how they interact with the CGM remains elusive, and a key challenge for modern galaxy formation theories (see \citealt{Somerville2015, Naab2017} for reviews).

Low-mass dwarf galaxies provide valuable clues to address these questions, as their shallower gravitational potential wells makes them acutely sensitive to their internal stellar processes. In particular, supernovae (SNe) explosions provide an important engine to inject energy and momentum into the ISM and launch powerful galactic winds capable of escaping such low-mass systems (e.g. \citealt{Chevalier1985, Dekel1986, Murray2005}). Despite this well-established picture, large uncertainties however remain in the quantitative efficiency with which these SN-driven outflows regulate star formation in dwarf galaxies.

Already within the ISM, additional stellar processes and feedback channels (e.g. winds, radiation; see \citealt{Naab2017} for a review) can modify the properties of the gas in which SNe explode, leading to qualitative and quantitative changes in the venting behaviour of low-mass galaxies (e.g. \citealt{Emerick2020StellarFeedback, Agertz2020EDGE, Smith2020PhotoRT, Fichtner2022}). Moreover, for a given feedback model, the spatial distribution of the energy injection and the clustering in space and time of SNe plays a major role in the ability to build a powerful outflow (e.g. \citealt{Girichidis2016SNOutflows, Kim2017, Fielding2018, Ohlin2019}), making outflow properties strongly sensitive to the underlying model for where young stars form and spatially distribute (e.g. \citealt{Andersson2020, Andersson2023Inferno,Steinwandel2023Runaways}). Further, the amount of feedback energy and momentum available from stellar evolution at the low metallicities of dwarf galaxies remains debated, making the total feedback budget itself an important uncertainty in such systems (e.g. \citealt{Gutcke2021Model, Prgomet2022}). All these factors contribute to the large spectrum of outflow energetics reported in the literature, with the predicted efficiency with which outflows carry mass and energy out of dwarf galaxies spreading over orders of magnitudes at a given stellar mass (see e.g. \citealt{Muratov2015, Hu2019, Emerick2020StellarFeedback, Smith2020PhotoRT, Pandya2021, Andersson2023Inferno, Steinwandel2022OutflowDwarf} for recent studies). 

Observational constraints are thus instrumental to help pinpoint the energetics of galactic outflows in low-mass galaxies, but the multiphase nature of such outflows makes this a challenging task. The different cold, warm and hot gas phases are probed through a diversity of observational techniques, each with their own wavelength window and challenges (see \citealt{Collins2022} for a review). Moreover, there is no consensus as to the respective importance of each phase in regulating star formation. The cold-to-warm phase ($\approx 10^4\, \K$) is expected to dominate the mass budget of the outflow (e.g. \citealt{Kim2018Tigress, Kim2020, Fielding2022}), but studies targeting this temperature range, i.e. through \halpha$\,$ or ultraviolet (UV) absorption lines, have both reported above-unity mass-loading factors (\citealt{Chisholm2017, McQuinn2019Outflows, Schroetter2019}) and much lower values (\citealt{Marasco2022LowOutflows}). Alternatively, the hot outflow phase ($\geq 10^{5}\, \K$) is expected to dominate the energy budget, which could be key to pressurize the surrounding CGM and prevent further star formation by delaying gas inflows onto the galaxy (e.g. \citealt{Hu2019, Li2020HotOutflows, Carr2023}). However, this phase is challenging to observe, emitting X-rays that are only accessible for limited samples of low-mass dwarfs (e.g. \citealt{Heckman1995, Summers2003, Summers2004, Ott2005, McQuinn2018}). This makes obtaining a holistic overview of outflow thermodynamics difficult observationally, and robust predictions of the multiphase structure of outflows paramount theoretically. 

\begin{figure*}
  \centering
    \includegraphics[width=\textwidth]{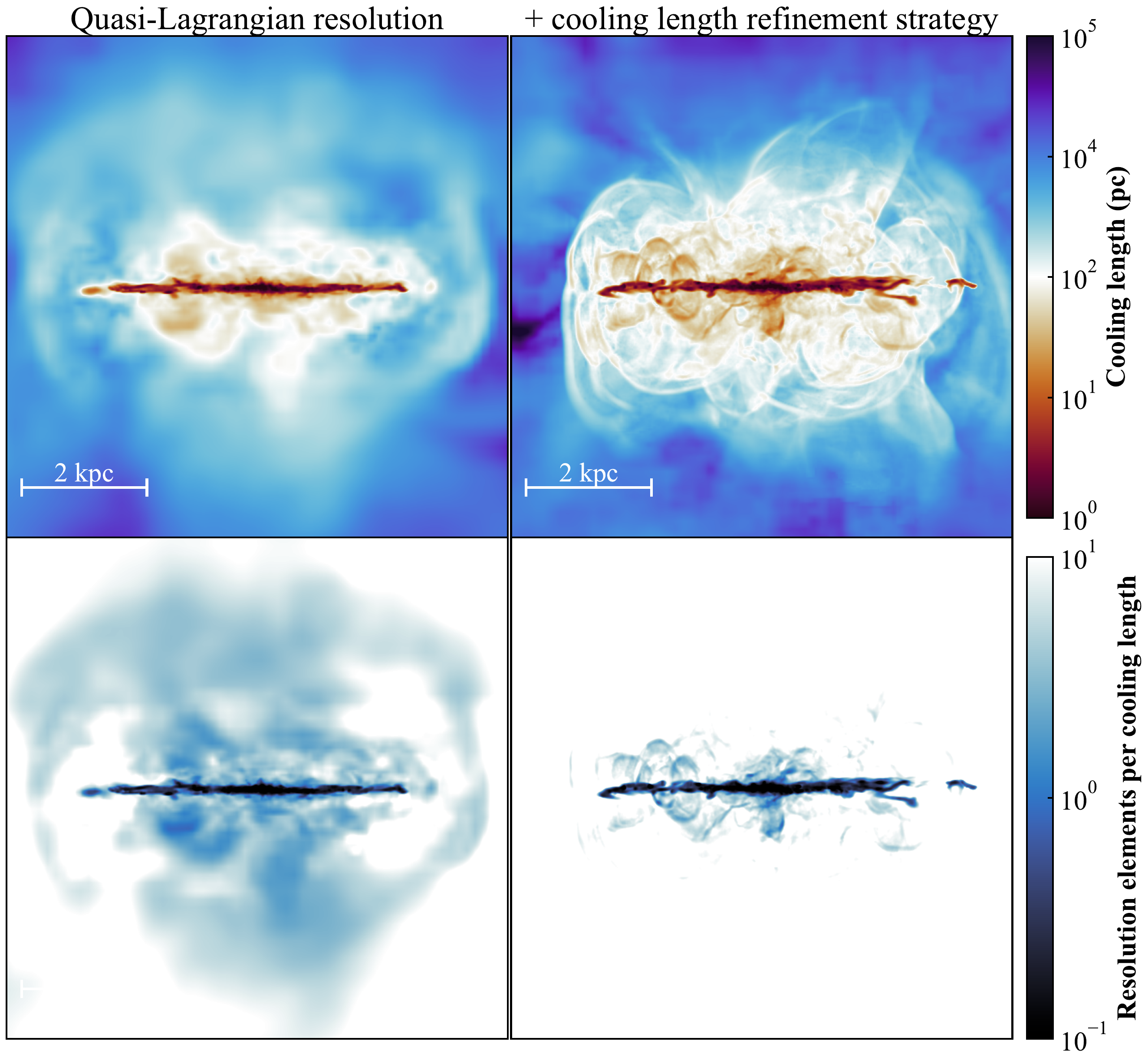}

    \caption{Visualization of the impact of better resolving the diffuse gas cooling length (right) compared to a traditional ISM-focussed quasi-Lagrangian strategy (left). Panels show edge-on maps of the gas cooling length (top) and its ratio with the spatial resolution of the simulation (bottom), density-weighted along the line of sight at a time of comparable outflow rates ($t = 540 \, \Myr$, $\approx 0.05 \, \Msolyr$). When resolution is focussed on the ISM (left), the dense outflowing material (top, orange to white, $\lcool \approx 10-100 \, \pc$) mixes rapidly into a diffuse hotter halo (blue, $\lcool \geq 100 \, \pc$). As the resolution degrades rapidly in these low densities (e.g. Figure~\ref{fig:amr}), cooling processes are marginally or significantly under-resolved (bottom left, blue). Forcing refinement on $\lcool$ (right, here down to $18 \, \pc$) resolves the thermodynamics of the outflowing gas much better, leaving only thin interfaces of under-resolved cooling gas outside the disc plane (bottom right). This improved treatment significantly enhances outflow energetics (Section~\ref{sec:outflowstructure}) and modifies their ionic structure (Section~\ref{sec:ionbyion}). Accurately capturing the launching and cooling of outflows within the dense ISM requires sub-pc resolution (top, dark red) posing a distinct, complementary challenge for galaxy formation simulations.
    }
    \label{fig:examplecoolinglength}
\end{figure*}

Beyond the launching mechanics of the wind and its initial structure within the ISM, a key aspect to achieving accurate predictions is to robustly capture the propagation of the multiphase outflow as it escapes the galaxy. A spectrum of processes can indeed efficiently transfer mass and energy between gas phases during the propagation, in particular shocks (e.g. \citealt{Klein1994}), hydrodynamical instabilities and turbulence (e.g. \citealt{Gronke2018, Gronke2020, Fielding2020, Kanjilal2021, Tan2021}), cooling and heating of the different gas phases (e.g. \citealt{Field1965, Begelman1990, Sharma2010, Sharma2012, McCourt2012, Voit2015}), the interaction with a surrounding hot or already multiphase CGM medium (e.g. \citealt{Armillotta2016, Bruggen2016, Gronke2022}), or a combination of the above (see \citealt{Faucher-Giguere2023} for a review). The relative efficiency of each of these processes remains debated as they depend on the local gas and radiation conditions, but they introduce characteristic length scales that should be resolved to accurately capture how the original structure of outflows is reprocessed as they expand away from the galaxy. A key common feature for these processes, however, is that their characteristic length-scales are invariably small ($\ll 10 \, \pc$), posing a huge numerical challenge to simulations that wish to model entire galaxies ($\geq \kpc$).

This issue is further compounded by the Lagrangian behaviour of most galaxy simulations, focussing the computational effort and resolution where gas is dense (i.e. in the ISM) but quickly degrading it in the diffuse gas to speed up the computation. Density and temperature contrasts in the launched outflow close to the galaxy are then numerically smoothed out as it expands away and the resolution worsens. This leads to artificial over-mixing of the gas phases and the suppression of thermal instabilities which we illustrate in Figure~\ref{fig:examplecoolinglength} (see also the discussion in \citealt{Hummels2019}). 

Figure~\ref{fig:examplecoolinglength} shows the gas cooling length density-weighted along the line of sight in the edge-on isolated dwarf galaxy used in this study. The cooling length is $\leq 100\, \pc$ in the outflowing gas escaping the galaxy (top, blue to white to red), but is marginally or under-resolved in a traditional galaxy formation simulation relying on a quasi-Lagrangian resolution strategy (bottom left, blue). The multiphase structure of simulated galactic outflows is thus likely to be under-resolved and numerically suppressed, reminiscent of the numerical limitations in modelling the larger-scale diffuse CGM (e.g. \citealt{Peeples2019, Hummels2019, Suresh2019, vandeVoort2019}). These challenges have motivated subgrid implementations and new hydrodynamical methods to model multiphase galactic winds (e.g. \citealt{Huang2020, Huang2022, Weinberger2023, Smith2024}), but a detailed understanding of how the energetics and observables of outflows depend on resolution remains lacking.

Furthermore, outflows are highly dynamic in nature, propagating at high velocities ($\gtrapprox 100 \, \kms$) into highly stratified media where the density and strength of the radiation field steadily decrease away from the galaxy. These quickly-evolving physical conditions and their low-densities make the cooling, chemical and ionic composition of outflows particularly prone to non-equilibrium effects during their propagation, as ionic recombination timescales can significantly differ from the gas cooling time in low-density environments (e.g. \citealt{Oppenheimer2013}, see also \citealt{Sarkar2022} for further radiative transfer effects in outflows).

To gain a better understanding of the importance of these physical processes, we create a new suite of galaxy-formation simulations aimed to provide a more robust modelling of galactic outflows. We perform isolated radiation-hydrodynamical simulations of a dwarf galaxy with the AMR code \textsc{ramses-rtz} (\citealt{Katz2022RTZ}). This allows us to (i) obtain an accurate account of non-equilibrium effects by solving the non-equilibrium chemistry of $\geq$60 ionic species coupled on-the-fly with the spatially- and time-varying radiation field, while (ii) following the development of galaxy-scale outflows self-consistently powered by stellar feedback within the disc. We further complement this setup by implementing a new AMR refinement strategy in \textsc{ramses-rtz} that explicitly aims to resolve the local gas cooling length (see also \citealt{Simons2020} for a similar approach in the CGM), enabling us to better resolve instabilities and mass transfer between gas phases during the outflow propagation (Figure~\ref{fig:examplecoolinglength}, right column). 

In this paper, we focus on understanding how gradually improving the resolution in galactic winds impact their energetics, and will dedicate companion papers quantifying how this improved treatment affects their emission lines properties (see also \citealt{Cameron2023ISMTe} for a similar approach with \textsc{ramses-rtz}). We describe the simulation suite and our numerical methods in Section~\ref{sec:setup}. We then show in Section~\ref{sec:outflowstructure:thermodynamics} that improving outflow resolution enhances the prevalence of the cold and hot gas phases, both of which exhibit systematically increasing energetics (Section~\ref{sec:outflowstructure:kinematics}). At fixed feedback budget, this results in a greater than five-fold increase in mass, energy and metal loading factors away from the galaxy (Section~\ref{sec:outflowstructure:loadingfactors}) and a doubling of high-ionization (e.g. \ovi) covering fractions (Section~\ref{sec:ionbyion}). We discuss our results and new approach in Section~\ref{sec:discussion} and conclude in Section~\ref{sec:conclusion}.

\section{Numerical setup} \label{sec:setup} 

We present a suite of radiative hydrodynamical simulations of isolated galaxies, all performed with the same physical modelling but gradually increasing off-the-plane spatial resolution to resolve thermal instabilities in the diffuse outflowing gas. We summarize our numerical galaxy formation model in Section~\ref{sec:setup:rtz}, in particular how we account for non-equilibrium, ion-by-ion cooling coupled to on-the-fly radiative transfer (see \citealt{Katz2022RTZ} for a more extensive description of the same setup). Section~\ref{sec:setup:refinement} presents the implementation of the new refinement strategy targeting the gas cooling length and Section~\ref{sec:setup:suite} the specific galaxy and simulations of this work.  

\begin{figure*}
  \centering
    \includegraphics[width=\textwidth]{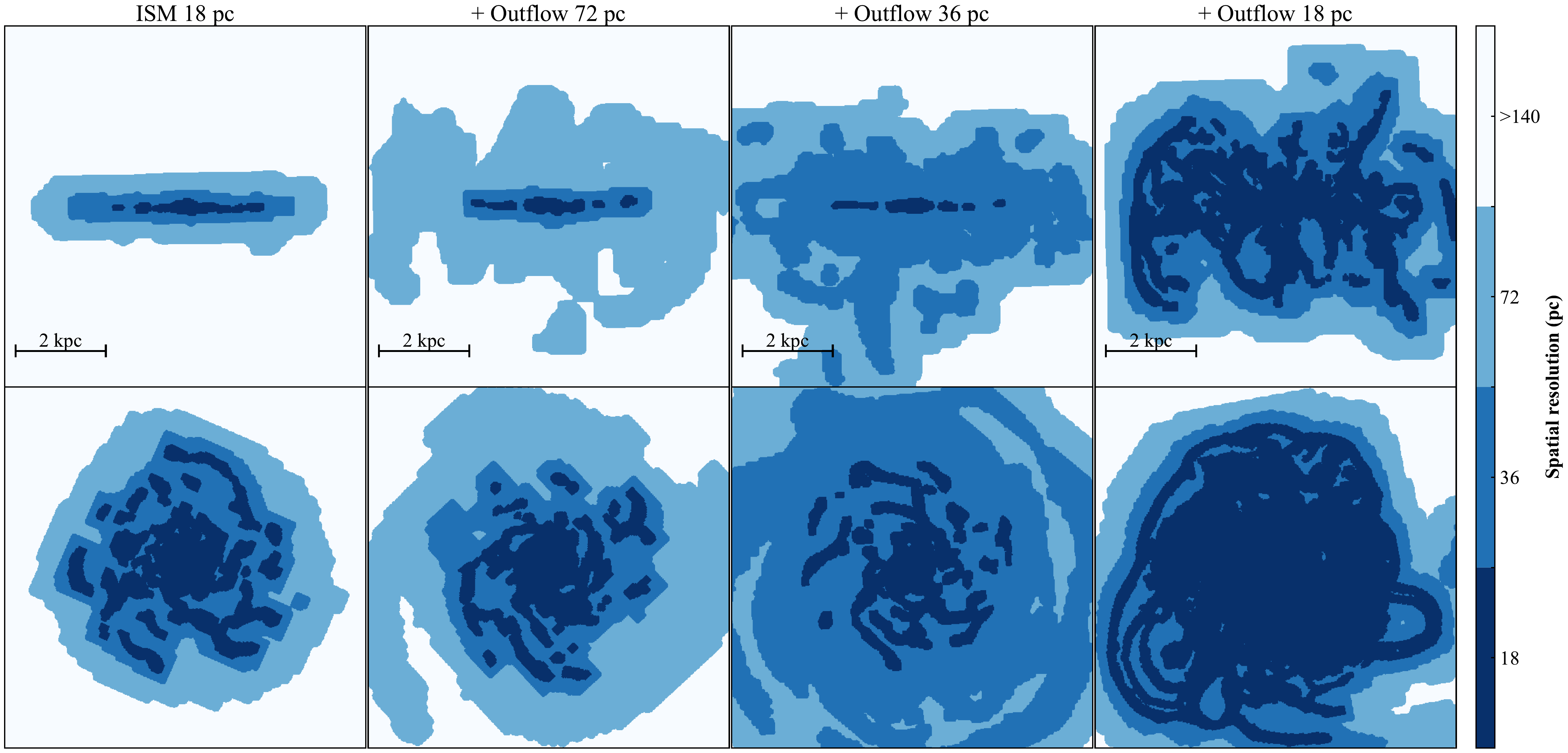}

    \caption{Edge-on (top) and face-on (bottom) AMR spatial resolution across a thin slice in the image plane, for the four galaxies with increasingly resolved $\lcool$ (left to right) at a time of comparable outflow rates ($t = 540 \, \Myr$, $\approx 0.05 \, \Msolyr$). The traditional refinement strategy (left) has a strongly layered vertical structure (top left) which under-resolves the off-the-plane cooling length (Figure~\ref{fig:examplecoolinglength}), transitioning quickly from the ISM resolution ($18 \, \pc$, deep blue) to $\geq 100 \, \pc$ (white). By contrast, resolution with our new scheme targeting the cooling length naturally tracks the interfaces of expanding superbubbles off the disc plane, following their irregular spatial structure and preventing them from numerically over-mixing. The structure of the dense ISM (bottom, deep blue) only gets visually modified when the additional refinement in the diffuse gas reaches a maximal resolution equal to that in the ISM (right-most). Nonetheless, star-formation conditions between all runs remain consistent with the fiducial setup, even for our most resolved cases (Appendix~\ref{app:similarsfs}).
    }
    \label{fig:amr}
\end{figure*}

\subsection{Chemistry, cooling and galaxy formation physics} \label{sec:setup:rtz}

We perform radiative hydrodynamical numerical simulations using the adaptive-mesh refinement code \textsc{ramses-rtz} (\citealt{Katz2022RTZ}), built upon the \textsc{ramses} and \textsc{ramses-rt} software (\citealt{Teyssier2002,Rosdahl2015RAMSESRT}). We solve the gas dynamics using a HLLC Riemann solver (\citealt{Toro1994}), assuming an ideal gas equation of state with adiabatic index of 5/3, while collisionless dynamics of stars and dark matter particles are computed by means of an adaptive particle-mesh method (\citealt{Guillet2011}).

A key novelty in our simulations is the treatment of ion non-equilibrium thermochemistry, coupled on-the-fly to the radiative transfer. We solve the dynamics of the radiation field at every timestep using the M1 method (\citealt{Rosdahl2013, Rosdahl2015RAMSESRT}), discretizing its spectrum in eight energy bins from the infrared to the UV (\citealt{Kimm2017}) and propagating radiation using the reduced-speed-of-light approximation ($c_{\text{reduced}} = c/100$). Radiation from local stellar sources is advected using the M1 method, while a uniform and fixed UV background also permeates the simulation box. Both sources contribute to photo-ionization and photo-heating, with the UV background following the tabulation of \citet{Haardt2012}. Gas above $\nh \geq 10^{-2} \, \cmcube$ self-shields exponentially from this UV background (\citealt{Aubert2010, Rosdahl2012}).

We then track the non-equilibrium evolution of 11 atomic species and 64 individual ions (H$~\textsc{i}-\textsc{ii}$, He$~\textsc{i}-\textsc{iii}$, C$~\textsc{i}-\textsc{vii}$, N$~\textsc{i}-\textsc{viii}$, O$~\textsc{i}-\textsc{ix}$, Ne$~\textsc{i}-\textsc{vii}$, Mg$~\textsc{i}-\textsc{vii}$, Si$~\textsc{i}-\textsc{vii}$, S$~\textsc{i}-\textsc{vii}$, Fe$~\textsc{i}-\textsc{vii}$ and Ca$~\textsc{i}$) and the formation and evolution of molecular hydrogen (\citealt{Katz2017}). Number densities of individual ions and species are computed at every simulation timestep by solving a network of coupled equations accounting for recombination, collisional ionization, charge exchange and photo-ionization extracted from the local, time-varying radiation field (\citealt{Katz2022RTZ}).

Having access to individual number densities of ions and the radiation field further allows us to self-consistently compute the out-of-equilibrium cooling rate for the gas. At low temperatures ($T \leq 10^4$ K), we explicitly compute the level populations and the fine-structure emission of \oi, \oiii, \ci, \cii, \siI, \siII, \sI, \feI, \feII, \neII$\,$ which dominate the cooling rate at the low metallicities we consider in this work (e.g. \citealt{Glover2007CoolingLowZ}). We use tabulated non-equilibrium ion-by-ion cooling tables for our UV background at high temperatures ($T \geq 10^4$ K; \citealt{Oppenheimer2013}) and further account for non-equilibrium atomic cooling processes of H and He (\citealt{Rosdahl2013}) and molecular cooling of $\hmol$ (\citealt{Katz2017}). 

This thermodynamical setup is complemented by an extensive galaxy formation model, including star formation and stellar feedback processes. Stars form in a molecular-cloud-like environment identified using a thermo-turbulent star-formation criterion (\citealt{Federrath2012, Kimm2017}) and with a \citet{Kroupa2001} initial mass function. Star particles are spawned with minimum initial masses of $1830 \, \Msol$. Stellar particles then re-inject energy, mass, momentum, metals and radiation locally, following the mechanical feedback prescription of \citet{Kimm2015} for SNe explosions and the model of \citet{Agertz2013} for stellar winds. Radiation is injected in gas cells surrounding stellar particles assuming a single stellar population of the age and metallicity of the particle and a BPASSv2.2.1 spectral energy distribution  (\citealt{Stanway2016}). The radiation impacts the gas via photo-ionization, radiation pressure, and photo-heating. These simulations model photo-electric and $\hmol$ heating following \citet{Katz2017} and ignore cosmic ray processes.
 
\subsection{Refinement strategy on the gas cooling length} \label{sec:setup:refinement}

All our adaptive-mesh simulations start by using a traditional `fiducial' quasi-Lagrangian refinement strategy targeting the ISM. To achieve this, we split AMR cells when their dark matter plus baryonic mass exceeds 8,000 $\Msol$ for 13 levels in a $150 \, \kpc$ box, corresponding to a maximum spatial resolution of $18$ pc. We also split cells when the local Jeans' length is resolved by less than four cells to obtain a more homogeneous spatial resolution across the ISM. This allows us to better capture ISM turbulence (e.g. \citealt{Federrath2011}) rather than solely focussing on dense clumps.

Figure~\ref{fig:amr} illustrates the structure of this fiducial refinement strategy (left-most column), showing an edge-on (top) and face-on (bottom) map of the spatial resolution at a time of comparable outflow rates between the simulations ($t = 540 \, \Myr$, $\approx 0.05 \, \Msolyr$). With the quasi-Lagrangian approach, resolution is near its allowed maximum within the disc ($18 \, \pc$, deep blue) but drops rapidly in the vertical direction ($\geq 100 \, \pc$, white) as the density decreases.

However, the cooling length of gas escaping the galaxy can be much smaller and is either marginally or under-resolved by the quickly degrading resolution (Figure~\ref{fig:examplecoolinglength}). To remedy to this issue, we store the net cooling rate $\Lambda_{\text{net}}$ obtained at each simulation timestep by the non-equilibrium solver and compute 
\begin{equation}
  \lcool = \sqrt{\frac{P_{\text{th}}}{\rho}} \times \frac{1}{t_{\text{cool}}} = \sqrt{\frac{P_{\text{th}}}{\rho}} \times \frac{3 \, \rho \, k_b \, T}{2 \, \mu \, \Lambda_{\text{net}}} \,   
  \label{eq:clength}
\end{equation} 
for each gas cell. Here, $P_{\text{th}}$ and $\rho$ are the thermal gas pressure and density used to define the isothermal sound speed. The cooling time, $t_{\text{cool}}$, is defined as the internal energy of the gas divided the net cooling rate $\Lambda_{\text{net}}$, with $\mu$ the mean molecular weight (mean mass per particle in units of the proton mass), and $\Lambda_{\text{net}}$ in units of $\text{erg} \, \text{s}^{-1}\, \cmcube$ (see \citealt{Sutherland1993}, eq 52 to 56 for  definitions). When computing $\lcool$ through Equation~\ref{eq:clength}, we use the self-consistent $\mu$, $\Lambda_{\text{net}}$, $T$ and $\rho$ computed internally by the non-equilibrium thermochemistry and hydrodynamical solvers. $\Lambda_{\text{net}}$ and $\lcool$ can be positive or negative depending on whether the gas is cooling or heating.

We then implement an additional refinement criterion, splitting gas cells in 8 where $\lcool$ is positive (i.e. only cooling gas) and unresolved by at least $\lcool / (N_{\text{cells}} \, \Delta x) < 1$, where $\Delta x$ is the spatial resolution at a given level of the mesh hierarchy. We allow $N_{\text{cells}}$ to take different values at different levels along the mesh hierarchy. 

The right columns of Figure~\ref{fig:amr} show the resulting spatial resolutions with this new strategy, resolving $\lcool$ with $N_{\text{cells}}=8$ down to 72, 36 and 18 pc (second, third and fourth columns respectively). The structure of the galactic disc (bottom) and its dense ISM (deep blue) where star-formation occurs is visually similar in all cases, until we reach a cooling length refinement matching the maximal ISM resolution (right-most column). Even in this case however, we show in Appendix~\ref{app:similarsfs}  that our new refinement scheme adds resolution in the diffuse gas around the disc, but does not impact the location and densities at which star formation occur and supernovae explode.

Rather, targeting $\lcool$ introduces large amount of additional off-the-disc-plane resolution that tracks discrete plumes and shells of outflowing gas as they expand and mix with the inner CGM (top panels of Figure~\ref{fig:amr}). The filamentary structure extending off the galaxy is visibly more extended from left to right, but remains contained by avoiding  unnecessary resolution in the hot halo where densities are lower and the cooling lengths larger (white). 

Since the ISM is already at the maximum resolution due to the quasi-Lagrangian scheme, the additional refinement criterion on the cooling length allows us to allocate additional resources to the diffuse gas in a controlled and gradual way. This provides us with the ability to cleanly and conveniently decouple the resolution between dense and diffuse gas, and to isolate their relative impact on outflow properties in Section~\ref{sec:outflowstructure}. 

Furthermore, we also verified that the resolution structure significantly varies with time and naturally follows to the star formation and outflow activity of the galaxy. This provides us with an efficient refinement strategy that focusses on the time-evolving physical quantity of interest rather than a pre-defined, arbitrary volume, and its adaptive nature also enables us to tailor the computational load to the application at hand. One can either model isolated outflows as resolved as the ISM (e.g. our most-resolved case in this study) or use more gradual strategies in the diffuse gas to limit the computational expense (e.g. our intermediate setups). We discuss the computational costs and pros and cons of different foreseen simulation strategies in Section~\ref{sec:discussion:cpucost}.

\subsection{Simulation suite} \label{sec:setup:suite}

All simulations in this work (Table~\ref{table:runs}) follow the evolution of the same, isolated dwarf galaxy, extensively described as G8 in \citet{Rosdahl2015RTFeedback, Kimm2018,Katz2022RTZ}. Briefly, we set up $2 \times 10^5$ particles to sample a dark matter halo of $10^{10} \, \Msol$, an initial disc and bulge (masses of $\xMsol{3}{8}$ and $\xMsol{3}{7}$, respectively) leading to a maximum circular velocity of $\approx 30 \, \kms$. Disc gas is initialized with an exponential density profile and a weak metallicity gradient peaking at $0.1 \, \Zsol$, and is surrounded by a homogeneous ($\rho=10^{-6}\, \mppercmcube$) hot, metal-free halo. We initialize all ions in their ground state in the disc and in their most ionized state in the hot halo.

\begin{figure}
  \centering
    \includegraphics[width=\columnwidth]{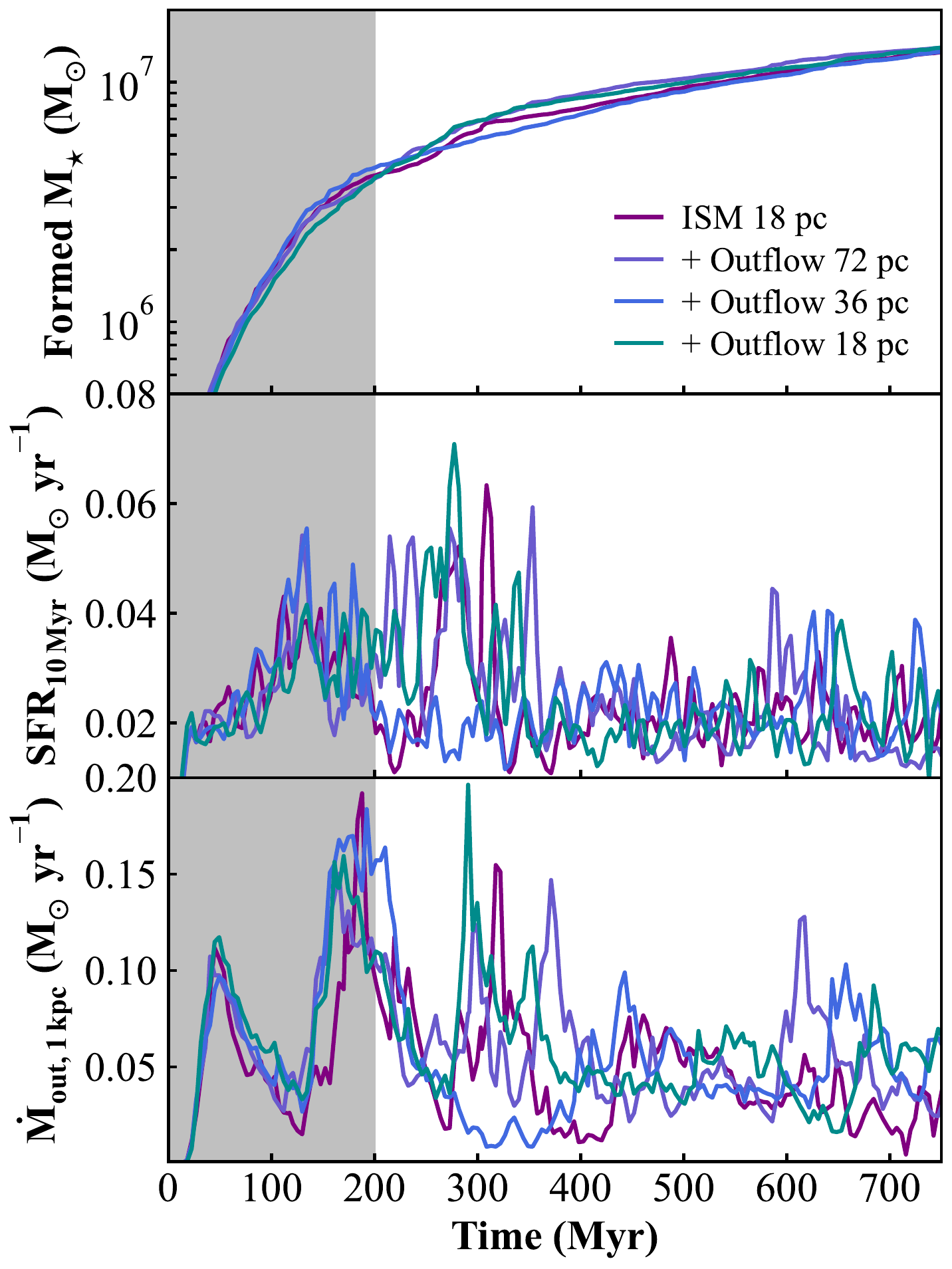}
    \caption{Stellar mass growth as a function of time (top), star formation rates averaged over 10 Myr (middle) and mass outflow rates at $|z| = 1\, \kpc$  (bottom) for our four galaxies with increasingly resolved outflows. Despite the change in numerical setup, all galaxies form a similar amount of stars, ensuring that the total feedback budget is approximately the same across all simulations. Better resolving the cooling length also yields comparable average star formation rates and mass outflow rates close to the disc, providing a controlled study. We discard the first 200 Myr of each simulation to allow the artificial transient phase induced by our initial conditions to dissipate.
    }
    \label{fig:timeevolution}
\end{figure}

\begin{figure*}
  \centering
    \includegraphics[width=\textwidth]{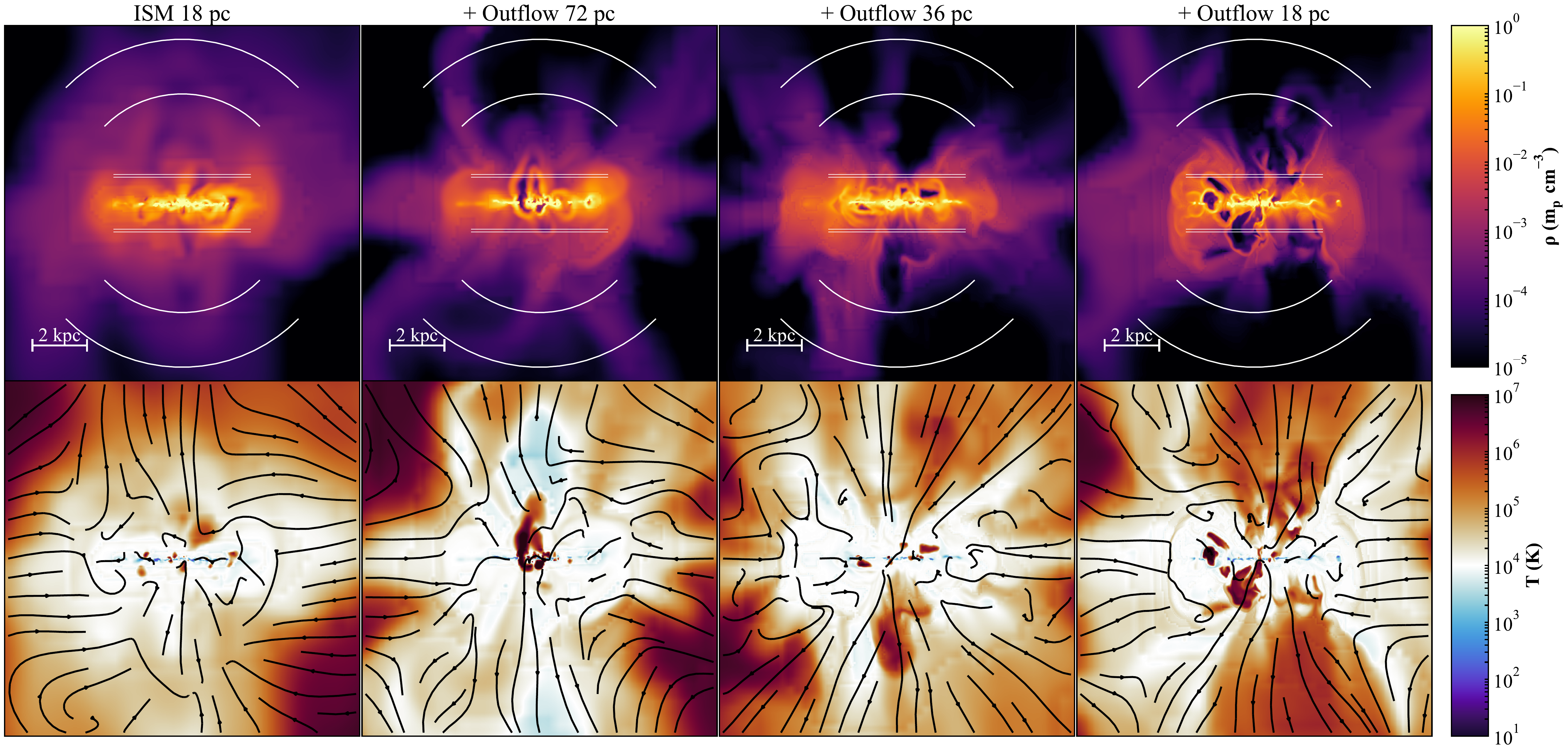}
    \caption{Thin slices through the gas density (top) and temperature (bottom) of the galaxy viewed side-on at the same time as Figure~\ref{fig:amr}. Increasingly resolving the cooling length of the diffuse gas (left to right) yields a visibly more structured outflow, better capturing the density and temperature contrasts between the hot, volume-filling cavities after shocks and the thin, dense shells of expanding material. At this time, velocity streamlines (bottom, black arrows) are also increasingly open, indicating a faster moving and more energetic outflow. Geometrical interfaces through which we quantify outflow energetics are visualized in the top panels (white lines).
    }
    \label{fig:fluidvisualization}
\end{figure*}

We perform four main simulations of this galaxy, progressively increasing the resolution target of the cooling length to 72, 36, and 18 pc eventually matching the ISM resolution. We follow the evolution of each galaxy for $750 \, \Myr$ (i.e. $\sim5$ disc crossing times) and save snapshots at least every 3 Myr. Figure~\ref{fig:timeevolution} shows the evolution of the stellar mass formed over the course of each simulation (top), their star-formation rates averaged over 10 Myr (middle) and mass outflow rates close to the disc of the galaxy (bottom, measured in a 0.1 kpc-thick slab placed at $|z| = 1\, \kpc$ and of cylindrical radial extent $R = 4 \, \kpc$; see Figure~\ref{fig:fluidvisualization} for a visualization). 

Our initial conditions drive an initial starburst which recedes after $\approx 100\, \Myr$, and it takes an additional $\approx 100\, \Myr$ for the galactic disc and outflow to settle after a large enhancement in mass outflow rate (grey shaded region in Figure~\ref{fig:timeevolution}). We thus start our analysis at $200 \, \Myr$, ensuring at least 130 snapshots of analysis when deriving time-averaged properties (Section~\ref{sec:outflowstructure}). We verified that our conclusions are unchanged if starting at $250 \, \Myr$ or $300 \, \Myr$.

\begin{table}
  \centering
  \caption{Summary of the simulations presented in this work (first column). We simulate the same initial conditions with a maximum resolution in the ISM of 18 pc, refining the outflows to an incrementally higher spatial resolution (second column). Two of our setups are simulated three times with different random seeds (first and third lines) to assess stochasticity in our results (Appendix~\ref{app:stochasticity}). We also provide a controlled, time-limited test of the computational cost of each setup (third column) discussed in Section~\ref{sec:discussion:cpucost}.} 

  \renewcommand{\arraystretch}{1.4}
\begin{tabular}{l c c c c c}
     \hline
      Simulation & Maximum cooling & Average timestep cost \\
       & length target (pc) & (CPU s / simulated Myr)\\

     \hline

     ISM 18pc (x3) & N/A & $1256^{+220}_{-247}$\\
     + Outflow 72pc & 72 & $1282^{+183}_{-242}$\\
     + Outflow 36pc (x3) & 36 &$1643^{+233}_{-267}$\\
     + Outflow 18pc & 18 &$4109^{+483}_{-473}$\\
     \hline
     \end{tabular}
   \label{table:runs}
\end{table}

All simulations form a similar amount of stellar mass, ending up within 15 per cent of each other. This is key to compare their outflows, as it guarantees that the overall energy, momentum and metal budget from stellar feedback is nearly unchanged between different resolution runs. It also provides further evidence that the physics of star formation in our simulations is not affected by the additional resolution in the diffuse gas (Appendix~\ref{app:similarsfs}).

However, the ability to drive galactic outflows also depends on the coupling between feedback and the surrounding gas, which in turn depends on the clustering of SNe and the density of the medium in which they explode. This is subject to stochasticity due to the probabilistic nature of our modelling of star formation (see \citealt{Genel2019, Keller2019} for further discussion). To quantify the magnitude of this scatter, we perform three simulations of our reference galaxy and three with the cooling length resolved down to 36 pc, solely varying the random seed for star formation in each of them. We show in Appendix~\ref{app:similarsfs} that star formation and feedback conditions within the ISM are comparable between both stochastic resimulations of a given setup and different resolutions of the cooling length. We further show in Appendix~\ref{app:stochasticity} that run-to-run stochasticity leads to different but statistically compatible outflow rates close to the disc, and that the induced scatter remains subdominant compared to differences in outflow properties and rates resulting from an increase in resolution when measured further away from the galaxy (Section~\ref{sec:outflowstructure}). This therefore confirms the causal association of differences in outflows with a more accurate treatment of their thermodynamics. 

\section{Energetics of better-resolved outflows} \label{sec:outflowstructure}

We start by visualizing the impact of better resolving the cooling outflow from the galaxy. Figure~\ref{fig:fluidvisualization} shows thin slices of gas density (top) and temperature (bottom) in a side-on view at the same time as in Figure~\ref{fig:amr}. As the cooling length of the diffuse gas is increasingly resolved (left to right), the thermodynamical structure of the outflow is visibly affected. Gas goes from a smooth and warm medium (left) to an increasingly structured distribution, with both colder and higher-density shells and hotter and lower density cavities (right). As we will see in Section~\ref{sec:outflowstructure:thermodynamics}, increased density contrasts and a more multiphase structure are a systematic consequence of the improved numerical treatment of the diffuse gas and extends beyond the specific time of Figure~\ref{fig:fluidvisualization}. Furthermore, in the bottom panels, we overlay velocity streamlines to visualize changes in gas kinematics. At this time, the streamlines also showcase greater opening angles with increasing resolution in the diffuse gas. Increased opening angles can boost the acceleration of galactic winds (e.g. discussion in \citealt{Martizzi2016}), although Appendix~\ref{app:openingangle} shows considerable variability over time in opening angles that makes systemic trends with numerical resolution  difficult to establish. We show in Section~\ref{sec:outflowstructure:kinematics} that better-resolved galactic outflows are indeed systematically faster moving and more energetic, significantly boosting outflow loading factors through the slab and shell visualized in Figure~\ref{fig:fluidvisualization} as a result (Section~\ref{sec:outflowstructure:loadingfactors}).

\begin{figure}
  \centering
    \includegraphics[width=\columnwidth]{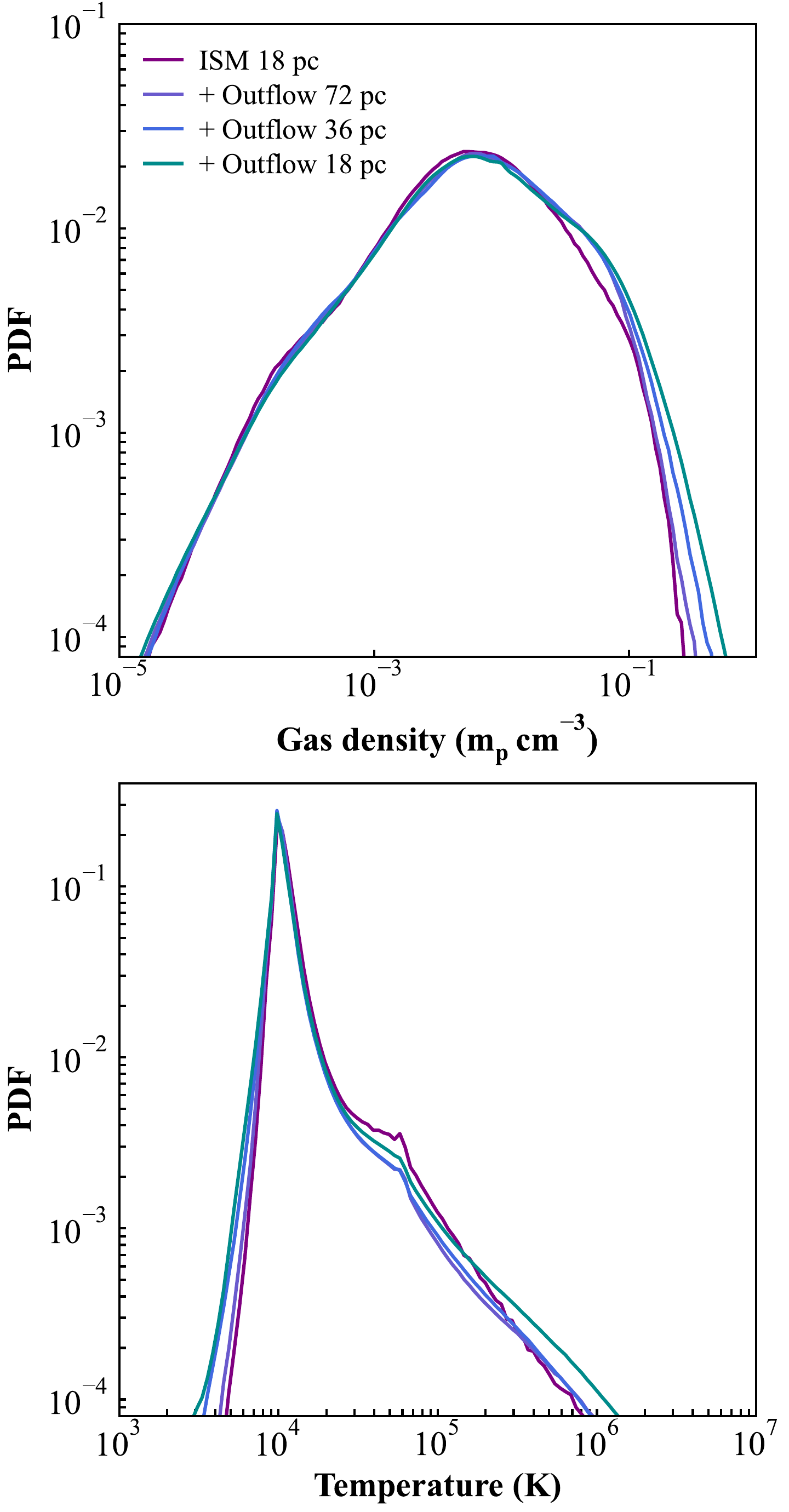}
    \caption{Time-averaged, mass-weighted distributions of density (top) and temperature (bottom) of the outflowing gas ($0.5 \leq z \leq 5\, \kpc$; $sgn(v_z) = sgn(z)$) across our suite of simulations. A better resolved cooling length (purple to green) leads to longer tails towards both denser and more diffuse gas, and both colder and hotter gas. This points towards an increasingly multiphase structure of galactic outflows with increasing resolution (Figure~\ref{fig:prhos}).  
    }
    \label{fig:trhospdfs}
\end{figure}

\begin{figure*}
  \centering
    \includegraphics[width=\textwidth]{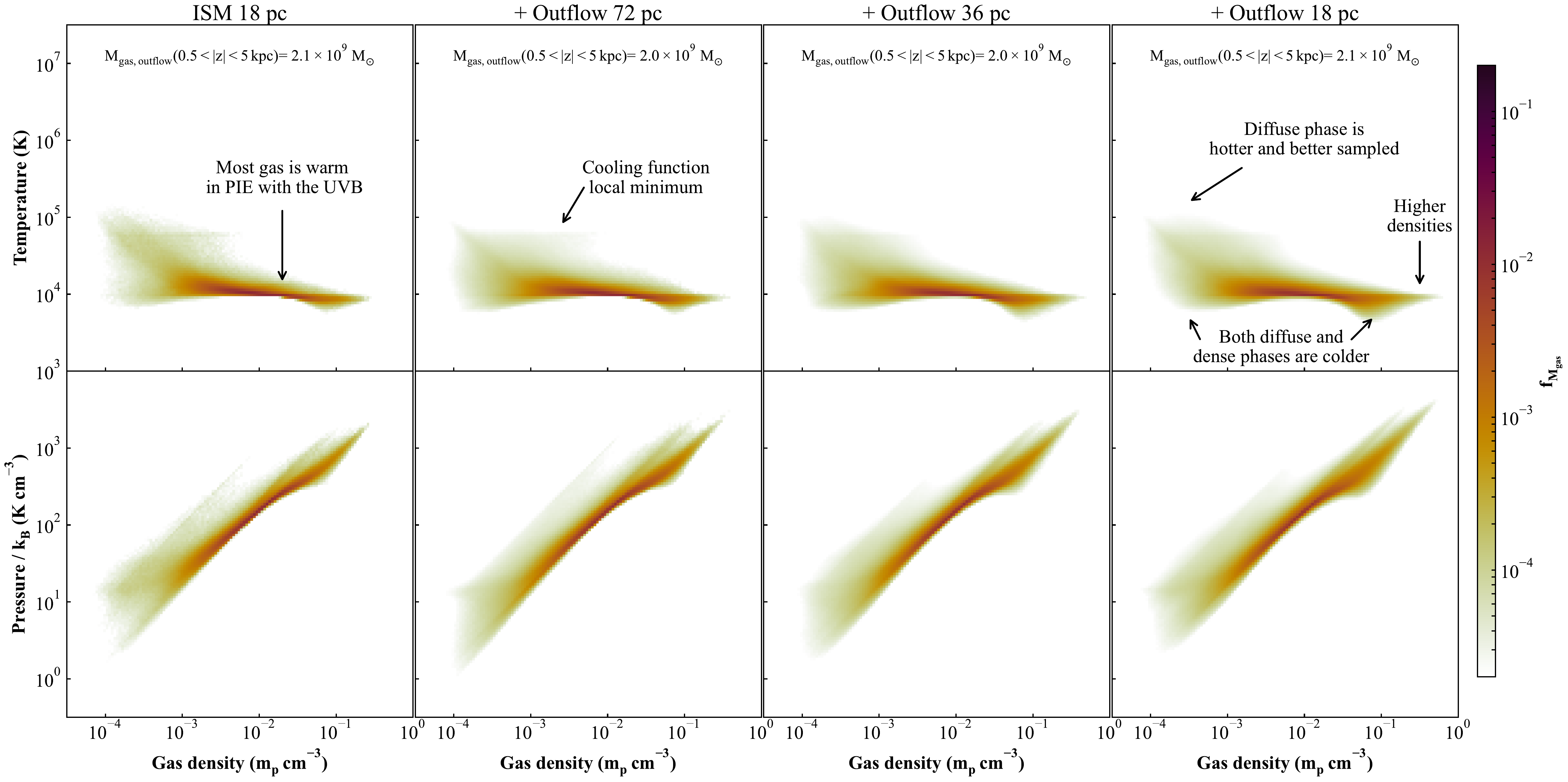}\\
    \caption{Time-averaged, mass-weighted temperature-density (top) and pressure-density (bottom) distributions of the outflowing gas when increasingly resolving the cooling length (left to right). In all cases, most gas mass is warm and in photo-ionization equilibrium with the UV background ($\approx10^4\, \K$). However, an increasing resolution yields a better-sampled diffuse phase with an increased scatter in temperatures and pressures at a given density, and allows denser and colder gas at high densities. This increasingly multiphase structure in turn affects how the thermo-kinematics of outflows as they propagate from close to the disc (Figure~\ref{fig:tvz_1kpc}) into the inner CGM (Figure~\ref{fig:tvz_5kpc}).
    }
    \label{fig:prhos}
\end{figure*}

\subsection{Increasingly multiphase outflows} \label{sec:outflowstructure:thermodynamics}
We first quantify the thermodynamical structure of our outflows over time. To this end, we extract a cylinder of gas above and below the disc plane ($R\leq 4.0 \, \kpc$ and $0.5 \leq |z| \leq 5 \, \kpc$) and select outflowing gas with $sgn(v_z) = sgn(z)$\footnote{We verified that relative trends reported here are conserved if instead} using $|z| > 1.0 \, \kpc$,  $|z| \leq 4 \, \kpc$, $|z| \leq 7 \, \kpc$ or $R\leq 5.0 \, \kpc$.. At each simulation snapshot, we then compute the mass-weighted two-dimensional histograms in pressure-density and temperature-density (150x150 pixels) of this gas. We stack these histograms pixel-by-pixel over $\geq 500 \, \Myr$ of time evolution, and normalize by the total gas mass to obtain the one- and two-dimensional PDFs in Figure~\ref{fig:trhospdfs} and Figure~\ref{fig:prhos}. Despite the change in resolution, all runs have similar total masses of outflowing gas (varying by less than 20 per cent, top of each panel in Figure~\ref{fig:prhos}), ensuring that changes in the PDFs are largely unaffected by changes in overall normalizations.  

Starting from the one-dimensional distributions for our fiducial case (Figure~\ref{fig:trhospdfs}, purple), we see that the majority of the outflowing gas mass is at low densities ($\sim 10^{-3} - 10^{-1} \, \mppercmcube$) and warm temperatures ($\sim 10^4\, \K$), as expected from diffuse gas in photo-ionization equilibrium with the surrounding UV background. A subdominant fraction of the gas is dense enough to self-shield efficiently against the UV background ($\rho \geq 10^{-1} \, \mppercmcube$ in our model) and cool below $10^4\, \K$. SN-heated gas is also visible as a long tail towards high temperatures ($\geq 10^{5}\, \K$), while the enhancement of gas with $T\approx \xScientific{8}{4}\, \K$ maps to a local minimum of the cooling function at the low metallicities considered in this work ($\leq 0.3 \, \Zsol$; e.g. \citealt{Bialy2019CoolingFunction, Katz2022RTZ, Kim2023}). 

Improving the numerical treatment of thermal instabilities (purple to green) does not modify this broad picture but introduces systematic trends for the tails of these distributions. The high-density and low-temperature tails incrementally reach higher densities and lower temperatures, confirming the expectation that condensation and cooling in the diffuse gas are numerically suppressed in our fiducial setup (Figure~\ref{fig:examplecoolinglength}). However, the improving resolution also generates a longer tail towards higher temperatures and lower densities, highlighting that the hot, diffuse phase is becoming increasingly significant. Since the total gas mass is roughly conserved between simulations, this is mainly modifying the relative fractions between gas phases, and thus the multiphase structure of our galactic outflows. 

To better visualize these changes, Figure~\ref{fig:prhos} shows the temperature-density (top) and pressure-density (bottom) two-dimensional phase diagrams of the same gas as in Figure~\ref{fig:trhospdfs}, recovering the broad features identified in the one-dimensional PDFs (top, annotated).

With increasing resolution (towards the right), the low-density gas ($\rho \leq 10^{-2} \, \mppercmcube$) is better sampled, generating a larger and smoother scatter in pressure and temperatures for a given density. This low-density gas also has a visibly cooler tail (top right), which we check arises from better-resolved ionized cavities following SNe explosions. These cavities cool from adiabatic expansion in a regime of inefficient recombination, preventing rapid heating from the UV background (see also \citealt{McQuinn2016} for a similar argument driving the temperature-density relation of the IGM). The dense gas ($\rho \geq 10^{-1} \, \mppercmcube$) cools more efficiently below $10^4\, \K$ with increasing resolution (top right), generating a more extended scatter in temperatures and starting to generate two distinct pressure tracks. 

Our results corroborate the physical picture in which the cooling length of diffuse outflowing gas from galaxies is under-resolved with traditional simulations (Figure~\ref{fig:examplecoolinglength} and Section~\ref{sec:intro}), hampering the development of multiphase galactic outflows. Improving resolution in the outflowing gas significantly helps alleviate this issue, yielding a more prominent colder and denser phase, but also a more prominent hotter and diffuse phase as we avoid over-mixing and over-cooling SN-driven expanding superbubbles. Both of these components are particularly important for galactic-scale feedback, as the cold-to-warm phase is expected to be significantly mass loaded, while the hot phase deposit energy in the surrounding diffuse CGM. We quantify the change in energetics of each of these gas phases in the next Section to gain better insights into these aspects.

Before turning to this, we briefly highlight that the one- or two-dimensional PDFs do not show signs of numerical convergence with increasing resolution and that the ordering of simulations with resolution is not linear (e.g. blue versus green in Figure~\ref{fig:trhospdfs}). This hints that, despite improving the situation, we are yet to fully resolve the multiphase structure of the outflowing gas and that evolution at a given resolution possesses stochasticity due to the specific star formation history at hand. We quantify this stochasticity in Appendix~\ref{app:stochasticity}, showing that it is a sub-dominant effect compared to the broad trends we identify when increasing resolution, and discuss remaining numerical uncertainties and convergence in Section~\ref{sec:discussion:convergence}.

\begin{figure*}
  \centering
    \includegraphics[width=\textwidth]{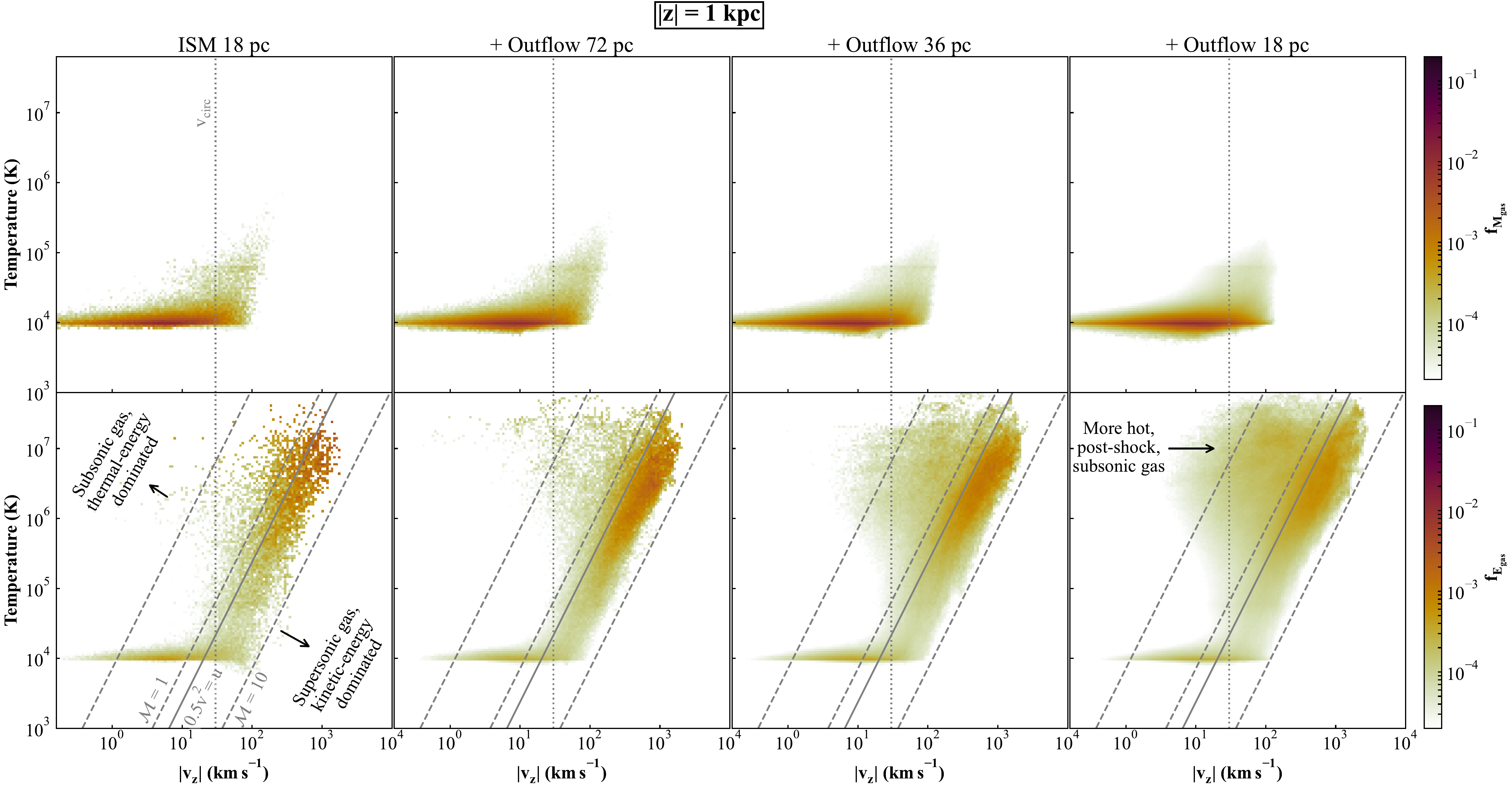}

    \caption{Time-averaged temperature-velocity distributions with increasing resolution (left to right), weighted by mass (top) and specific energy (bottom) for gas within a 0.1-kpc thick slab at $|z|=1\, \kpc$. Most outflow mass (top panels) is warm and moving slower than the escape velocity of the system (grey dotted in top panels). But better-resolved outflows are systematically more energetic (bottom), extending towards hotter temperatures and faster moving velocities (top-right corner of each bottom panel) along a track of near-constant Mach number (grey dashed). Each level of increased resolution also makes the cloud of hot and subsonic gas more prevalent, indicating that we increasingly resolve the sonic transition of the hot phase. This is key to accurately capturing the acceleration of the hot gas and its subsequent energetics as it escapes into the CGM (Figure~\ref{fig:tvz_5kpc}).
    }
    \label{fig:tvz_1kpc}
\end{figure*}

\begin{figure*}
  \centering
    \includegraphics[width=\textwidth]{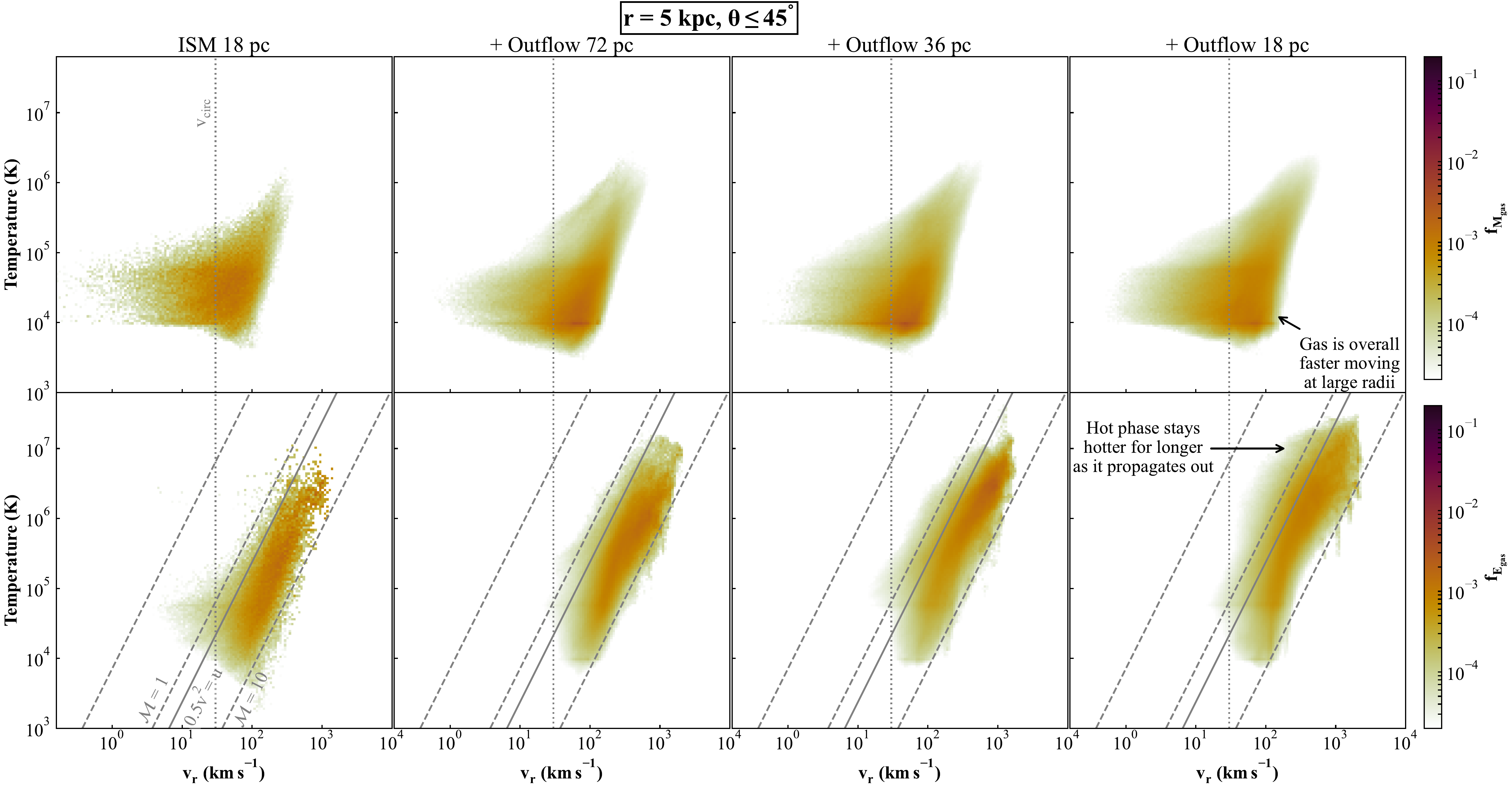}

    \caption{Same as Figure~\ref{fig:tvz_1kpc} but for gas within a 2-kpc wide shell at $r=5\, \kpc$ with opening angle $\theta=45^{\circ}$ along the minor axis of the galaxy (see Figure~\ref{fig:fluidvisualization} for a visualization). At these outer radii, outflow mass is hotter and faster moving ($v\geq 10\, \kms$) than at $|z| = 1\, \kpc$. The improved numerical treatment of the thermo-kinematics of the gas significantly boosts its energetics, with outflow mass systematically travelling at faster velocities (top panels) and a more energy-loaded hot-fast phase (bottom panels). The combination of these factors translate into large boosts in outflow loading factors (Figure~\ref{fig:loadingfactors}).
    }
    \label{fig:tvz_5kpc}
\end{figure*}

\subsection{Kinematics and energetics of better resolved outflows} \label{sec:outflowstructure:kinematics}

To gain physical insights into these differences in outflow energetics, we show in Figure~\ref{fig:tvz_1kpc} the two-dimensional distributions of temperature and vertical velocities of outflowing gas close to the disc (a 0.1-kpc slab at $|z| = 1 \kpc$ with $sgn(v_z) = sgn(z)$). The histograms are weighted by gas mass and total gas specific energy (defined as the sum of the kinetic and internal specific energies, $e_{\text{gas}} = 1/2 \, \ v^2 + u = 1/2 \, \ v^2  + P_{\text{th}} / \rho / (\gamma - 1)$) in the top and bottom panels, respectively. Again, each panel shows the normalized stack over the time evolution of each galaxy, with increasingly resolved cooling length from left to right.

From Figure~\ref{fig:tvz_1kpc}, we recover that most of the mass-weighted outflow gas (top panels) is warm ($\approx 10^4\, \K$), travelling at relatively low velocities ($\vz \leq 10 \, \kms$) and is unlikely to escape the gravitational potential of the galaxy ($\vcirc$ as grey dotted line). This picture remains qualitatively similar as the cooling length is increasingly resolved (left to right), although the improved resolution leads to a tail of $\leq 10^4\, K$ gas (top, left panel). This confirms that, close to the disc, the ejection of mass by SNe-driven outflows is not significantly modified by adding additional resolution in the diffuse gas (recall Figure~\ref{fig:timeevolution} and see also Appendix~\ref{app:similarsfs}). 

In stark contrast, the improved numerical treatment of the hot phase at higher resolution yields significant differences when weighting the same histograms by gas specific energy (bottom panels). Focussing first on the fiducial case (left), the hot, fast-moving and energy-carrying gas phase of the outflow materializes as a track extending towards high temperatures and velocities, broadly following an iso-Mach slope (dashed lines, where $\mathcal{M} = v / c_s$ with $c_s = \sqrt{P_{\text{th}} / \rho}$ the isothermal sound speed). This `plume' arises from the superposition of the many SN-driven superbubbles expanding out of the disc plane (see also \citealt{Kim2020, Andersson2023Inferno, Steinwandel2022OutflowDwarf}). This gas is strongly supersonic, and dominated by its kinetic energy (solid line shows the equality between gas kinetic and internal energy). Gas sampling the radiative regions behind shocks (hot and subsonic, towards the top left) is present, but poorly sampled in the fiducial case.

As we increasingly resolve the cooling length in diffuse gas (left to right), the plume of hot and fast-moving gas extends towards faster velocities and higher temperatures (respective upper-right corners of each panel). The hot phase is thus launched more energetically close to the disc. Furthermore, a significant hot, but subsonic, phase becomes ever more apparent with each level of increased resolution in the cooling length. This is to be expected as, by design, our approach concentrate resolution in rapidly cooling regions such as the post-shock, radiative layers that generate hot and fairly slow-moving ($\vz \approx 50 \, \kms$) gas. Combining these aspects, we conclude that our improved resolution scheme drastically improves the treatment of the thermo-kinematics of the hot phase by better capturing the expansion and cooling around shocked regions and keeping gas hotter for longer by avoiding spurious mixing and numerically-driven rapid cooling. 

We can expect such improvements at $z=1\, \kpc$ to be reflected in the energetics of the wind at larger radii, as better capturing the transition of the hot phase from sonic to supersonic is key to capture the global acceleration of galactic outflows (e.g. \citealt{Chevalier1985, Smith2024}). To visualize this, we repeat the same computations as for Figure~\ref{fig:tvz_1kpc}, but now selecting gas at $r=5\, \kpc$ ($\approx 0.2 \, \rvir$) and plotting $v_r$-$T$ to quantify differences in outflow thermo-kinematics after propagation into the inner CGM. In Figure~\ref{fig:tvz_5kpc}, we only include gas within a biconical shell with a broad opening angle $\theta = 45^\circ$ to capture the opening of the velocity streamlines. We show in Appendix~\ref{app:fullshell} that this selection removes gas from polar angles close to the plane of the disc that is often dominated by the hot hydrostatic halo remaining from the initial conditions of our galaxy (rather than the outflowing gas of interest).

At $r=5\, \kpc$, the mass-weighted outflow (Figure~\ref{fig:tvz_5kpc}, top panels) is hotter and faster moving than at $z=1\, \kpc$, as expected if the gas is accelerating as it escapes the galaxy. Furthermore, marginalizing the mass-weighted diagrams of Figure~\ref{fig:tvz_1kpc} and~\ref{fig:tvz_5kpc} along the y-axis, we then compute the mass-weighted velocity PDF. We find that both these PDFs shifts towards larger velocities with increasing resolution, but the increase is relatively mild at $z=1\, \kpc$ (medians going from 6.7 to 8.5 to 8.5 to 9.2 $\kms$ with each added resolution level, respectively) and more significant at $5\, \kpc$ (going from 39.9 to 63.4 to 46.5 to 50.3 $\kms$). In all cases, the outflowing gas significantly accelerates between $z=1\, \kpc$ and $r=5\, \kpc$, but this acceleration is amplified with higher resolution outflows.  

Weighting the outflow by energy confirms this picture (bottom panels in Figure~\ref{fig:tvz_5kpc}). Even after propagating to $r=5\, \kpc$, the hot phase remains better resolved with our improved scheme, extending towards hotter temperatures and faster velocities as expected from better capturing hot-to-warm interfaces (see e.g. \citealt{Kim2018Tigress} for a discussion). At our highest resolution (right column), we even recover a small hot, subsonic component (although much less prominent than at $z=1\, \kpc$) that is entirely absent with the fiducial resolution scheme. This highlights that we now capture the thermodynamics of shocks and the sonic transitions of specific superbubbles in the wind out to large radii. As we will see in the next Section, this results in significant boosts to the outflow loading factors with resolution.

\subsection{Consequences on outflow loading factors} \label{sec:outflowstructure:loadingfactors}
  
\begin{figure*}
  \centering
    \includegraphics[width=\textwidth]{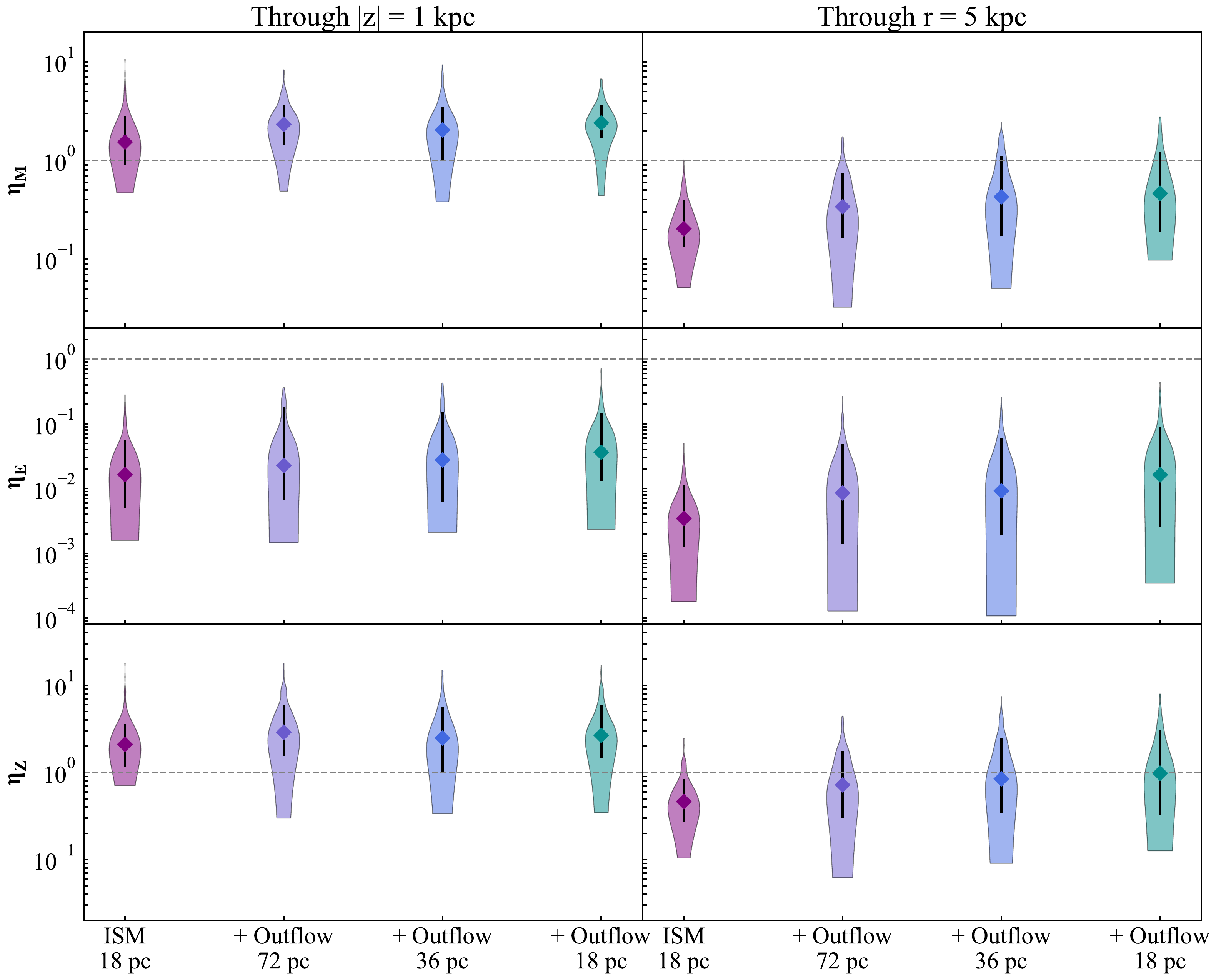}

    \caption{Mass, energy and metal loading factors (top, middle and bottom rows, respectively) measured through a slab close to the disc ($|z| = 1\, \kpc$, $R=4.0 \, \kpc$; left) and a biconical shell at large radii ($r = 5 \, \kpc$, $\theta \leq 45^{\circ}$; right). Violins show the distribution of loading factors over time as the star formation fluctuate during each galaxy's evolution (median in diamonds and 16-84 confidence interval as black line). Better resolving the dynamics of multiphase outflows from the low resolution fiducial case (purple) to the highest resolution case (green) systematically increases all loading factors. Close to the disc (left), increases are limited for the mass and metal loading factors, but grow to a two-fold increase at larger radii (right). The better-captured propagation of the hot phase significantly enhances the energy loading factors through both interfaces, culminating in a $5 \times$ boost at 5 kpc.  
    }
    \label{fig:loadingfactors}
\end{figure*}

We now turn to quantifying the impact of the improved numerical treatment of the outflowing gas on its loading factors. We compute, at each saved simulated time, the mass, energy and metal loading factors as
\begin{equation}
  \begin{split}    
    \eta_{M} &= \frac{\dot{M}_{\text{out}}}{\SFR_{\text{10 Myr}}} = \frac{\sum_{i \in \mathcal{S}} \, m_i \, v_i}{\SFR_{\text{10 Myr}} \ \Delta \mathcal{S}}\, ,\\
    \eta_{E} &= \frac{\dot{E}_{\text{out}}}{\SFR_{\text{10 Myr}} \, e_{\text{SN}}} = \frac{\sum_{i \in \mathcal{S}} \, (1/2 \, v_i^2 \, + \, \gamma u_i) \, m_i \, v_i}{\SFR_{\text{10 Myr}} \, e_{\text{SN}} \ \Delta \mathcal{S}} \\
    \eta_{Z} &= \frac{\dot{Z}_{\text{out}}}{\SFR_{\text{10 Myr}} \, Z_{\text{disc}}} = \frac{\sum_{i \in \mathcal{S}} \, m_i \, v_i \, Z_i}{\SFR_{\text{10 Myr}} \, Z_{\text{disc}} \ \Delta \mathcal{S}}.
  \end{split}
  \label{eq:loadingfactors}
\end{equation}
Here, $\mathcal{S}$ is the interface with thickness $\Delta \mathcal{S}$ through which the outflow rate is defined (respectively, a 0.1-kpc thick horizontal slab placed 1 kpc above and below the disc, and a 2-kpc thick shell placed at $r=5\, \kpc$ and polar angle $\theta \leq 45^\circ$; see Figure~\ref{fig:fluidvisualization}). $m_i$, $u_i$, $v_i$ and $Z_i$ are the mass, specific internal energies, velocities and total metal mass fraction of individual gas cells within $\mathcal{S}$. We use $v_i = v_{z,i}$ in Equation~\ref{eq:loadingfactors} for the slab and $v_i = v_{r,i}$ for the shell to account for the opening of the streamlines (Figure~\ref{fig:fluidvisualization}). We normalize the outflow loading factors with the star formation rate of the galaxy averaged over the last $10\, \Myr$\footnote{Our conclusions are unchanged with a longer $50\, \Myr$-baseline that covers the full time window of SNe energy injection from a star cluster.}, the average specific energy injected by SNe for a \citet{Kroupa2002} mass function, $e_{\text{SN}} = \xScientific{5.2}{5} \, \text{km}^2 \, \text{s}^{-2}$ (\citealt{Kim2018Tigress}) and the average mass-weighted metal mass fraction within the disc $Z_{\text{disc}}$ (geometrically defined as gas with $R\leq 4.0 \, \kpc$ and $|z| \leq 0.5 \, \kpc$). Relative trends between outflow loading factors are unchanged if measuring outflow rates at $|z| = 0.5 \, \kpc$ or using other radial definitions (see Appendix~\ref{app:fullshell}).

Figure~\ref{fig:loadingfactors} shows the mass, energy and metal outflow rates (top, middle and bottom panels respectively) at 1 and 5 kpc (left and right panels respectively). Violin plots showcase distributions across the time evolution of each simulation, highlighting their medians (diamonds) and 16-84 confidence intervals (black lines). 

Outflows close to the disc (left panels) are significantly mass and metal loaded ($\massloading, \metalloading \geq 1$) but not energy loaded ($\energyloading \ll 1$). Enhancing off-the-disc-plane resolution leads to a mild but systematic evolution, with a 50 per cent enhancement in the median mass loading factor between the fiducial and the most resolved case, a factor of two growth in the energy loading factor, and a 20 per cent increase in the median metal loading factor. These limited increases in mass and metal loading are consistent with the relatively unchanged mass-weighted PDF in Figure~\ref{fig:tvz_1kpc} (top panels), while the increase in energy loading factor is consistent with the hotter hot phase and better captured transition of the sonic-to-subsonic transition close to the disc (Figure~\ref{fig:tvz_1kpc}, bottom panels). However, except for the energy loading factor, the significance of these increases remains difficult to interpret, as stochasticity due to varying star formation activity at different times (black lines show the 16-84 confidence intervals) as well as the overall noise in the evolution of a numerical setup at a given resolution (Appendix~\ref{app:stochasticity}) can generate shifts of a similar magnitude.

In contrast, differences further away from the galaxy (right panels) are far more significant, with a two-fold increase in mass loading (median $\massloading = 0.20, 0.34, 0.43, 0.47$ from purple to green) and metal loading factors (median $\metalloading = 0.46, 0.72, 0.84, 0.98$), and a five-fold increase in energy loading factor (median $\energyloading = 0.0034, 0.0085, 0.0090, 0.017$). Importantly, these shifts with enhanced resolution are close to exceeding the 16-84 confidence interval span of the fiducial case and are robust against run-to-run stochasticity (Appendix~\ref{app:stochasticity}). 

Our study shows that better resolving the diffuse outflowing gas provides a significant boost to their energetics and, in turn, to their ability to regulate galactic star formation. When escaping into the CGM, our increasingly-resolved outflows go from being marginally ($\eta \leq 1$) to significantly ($\eta \geq 1$) mass and metal loaded. This ability to maintain the high mass and metal loadings acquired in the ISM over super-galactic scales could have strong consequences for galactic outflows' ability to efficiently transport mass outwards and pollute the CGM and IGM with metals. Similarly, the growth towards well energy-loaded outflows at 5 kpc would naturally help pressurize the inner CGM and prevent further gas accretion onto the galaxy.

The full reach of these results remains to be determined in a cosmological context, however. Our outflows currently propagate in a low-density, static hot halo surrounding our idealized dwarf galaxy, rather than the layered, dynamic and inflowing CGM of a cosmological galaxy. However, their significance strongly motivates future studies quantifying how the gas cycle of galaxies is affected when outflows and inflows are better resolved.

\section{Ionic structure of better-resolved outflows} \label{sec:ionbyion}

\begin{figure*}
  \centering
    \includegraphics[width=\textwidth]{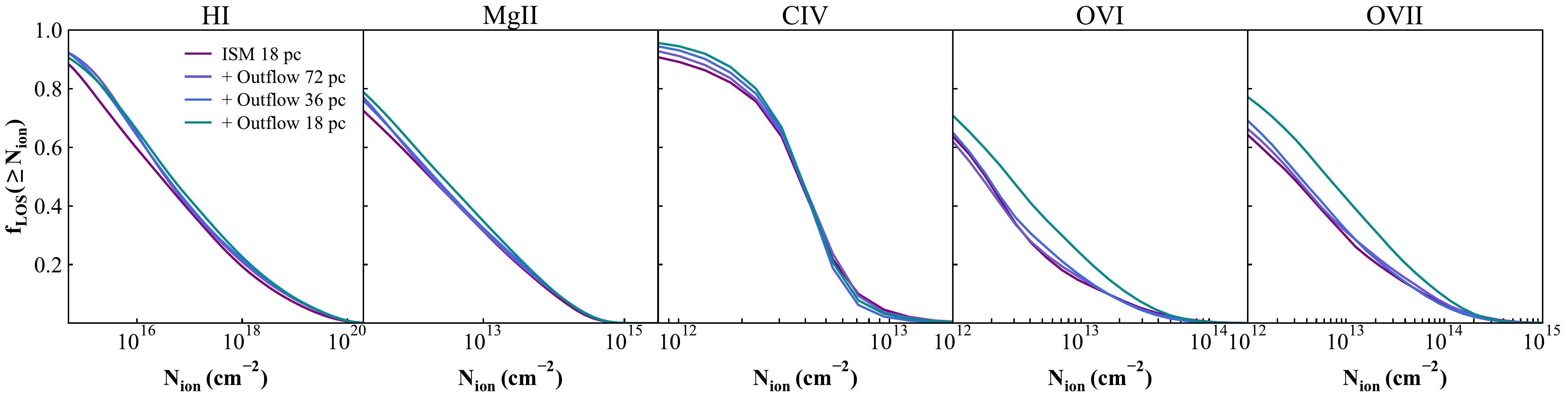}
    \caption{Time-averaged covering fractions of selected ions with increasing ionization potentials (left to right panels respectively) when looking at simulated galaxies edge-on in runs with increasing off-the-plane resolution (from purple for the fiducial case to green for the most resolved: see text for detail). A better resolved warm-to-cold phase leads to an increase in \hi and \mgII, particularly at low column densities ($\leq 10^{17} \, \cmsquare$ and $\leq 10^{12} \, \cmsquare$). A better resolved hot phase strongly enhances the covering fractions of high-ionization ions, notably in \civ and \ovi, two key ions to probe CGM gas in absorption.
    }
    \label{fig:coveringfractions}
\end{figure*}

A key novel aspect of our simulations is to track the non-equilibrium abundances of ions, providing us with the opportunity to readily study the ionic structure of our outflows and eventually link to emission and absorption observables. We now provide preliminary insights on the observational consequences of our findings, and will study specific observational diagnoses to recover gas properties from outflow observations in dedicated companion papers (see Section~\ref{sec:conclusion} for further discussion).

Figure~\ref{fig:coveringfractions} shows the time-averaged covering fractions of five ions with increasing ionization potentials. To derive those, we generate side-on images of the galaxy at each time output (8 kpc wide and deep, spatial resolution of 20 pc) masking the galactic disc ($|z| < 0.5 \, \kpc$) and compute histograms of line-of-sight column densities for each ion. We then aggregate the histograms over time, renormalize them to obtain the stacked  probability distribution functions, and take the cumulative sums shown in Figure~\ref{fig:coveringfractions}.

Starting from low-ionization ions (left panel), we observe a systematic, although mild, trend with resolution, with both $f({N_{\text{\hi}} \geq 10^{15}\, \cmsquare})$ and $f({N_{\text{\mgII}} \geq 10^{11}\, \cmsquare})$ increasing between our two extreme resolutions by 5 and 8 per cent, respectively. This is primarily driven by an increase at high column densities (e.g. $f({N_{\text{\hi}} \geq 10^{18}\, \cmsquare})$ increases by 15 per cent), which is best explained by the growth of a colder and denser gas phase as the out-of-the-plane cooling length is better resolved (Figure~\ref{fig:prhos}). The relatively limited evolution in the covering fractions of these low-ionization ions with resolution is encouraging, although the lack of convergence highlights that we are not yet resolving the characteristic scales regulating the dense, cold phase (see e.g. \citealt{vandeVoort2019, Gronke2020} and Section~\ref{sec:discussion:convergence} for further discussion). 

For ions tracking warmer, more diffuse and ionized gas (e.g. \civ, middle panel), we find a reduction in high-column density sightlines and a growth at low column densities with increasing resolution. We attribute the latter to the larger scatter of temperatures in the diffuse phase (Figure~\ref{fig:prhos}) yielding a more extended scatter in diffuse \civ$\,$  sightlines. In contrast, higher-densities sightlines hosting \civ$\,$ at low resolution have likely recombined and cooled to lower-ionization ions when the cooling length is better resolved, explaining the reduction. 

A much larger effect is visible when inspecting the covering fractions of higher ionization ions (right, \ovi, \ovii), with $f({N_{\text{\ovi}} \geq 10^{13}\, \cmsquare})$ going from to 0.148 to 0.160 to 0.169 to 0.246 with increasing resolution (0.306 to 0.321 to 0.327 to 0.437 for \ovii). These 66 and 43 per cent increases materialize the increasing energetics of our better-resolved outflows, with the systematically hotter gas promoting higher-ionization ions. 

The trend with resolution is however non-linear and difficult to interpret quantitatively. Lower-resolution runs (violet and blue) show mild evolution until a sudden increase in the covering fraction occurs with our most-resolved run. We checked that this is in part due to stochasticity in the given realization of the star formation and outflow history. For example, our two other realizations of the outflow at 36 pc (Appendix~\ref{app:stochasticity}) show 74 and 24 per cent increase in $f({N_{\text{\ovi}} \geq 10^{13}\, \cmsquare})$ compared to our fiducial simulation, while our other two random realizations of the fiducial setup vary by as much 27 per cent. Despite these quantitative uncertainties, we find that runs with improved treatment of thermal instabilities in the diffuse gas always have elevated covering fractions of \ovi$\,$ and \ovii.

Our results show that improving the treatment of thermal instabilities in the diffuse gas provides a significant avenue to increasing the covering fraction of high ionization ions. Probing the thermodynamics of the gas surrounding dwarf galaxies using ionic absorption lines is already possible (e.g. \citealt{Bordoloi2014, Johnson2017, Zheng2019,Zheng2020, Zheng2024, Qu2022DwarfCGM}), although we stress that our results should be compared with great care (if at all) to such data. Our idealized simulations lack a self-consistent large-scale CGM surrounding the dwarf, strongly limiting comparisons at large impact parameters and against statistical samples of dwarf galaxies. We verified that our results are qualitatively unchanged if including slightly higher impact parameters (using 10 kpc-wide images rather than 8), but refrain from detailed comparisons with data. 

Nonetheless, our results stress the importance of improving out-of-the-plane resolution to properly capture the ionic structure of outflows and constrain galaxy formation models from interpreting their observables in emission and absorption. These results further motivate applications of our setup and refinement strategy in a cosmological context to provide more robust comparisons against observational measurements at larger impact parameters, well into the CGM.

\section{Discussion} \label{sec:discussion}

\subsection{The numerical cost of refining on the cooling length} \label{sec:discussion:cpucost}

To establish that better resolving the multiphase structure of galactic outflows boosts their energetics, we rely on adding resolution in the diffuse thermally-unstable gas. This carries an associated numerical cost that potentially limits the practical scope of our approach and the applicability of our refinement strategy. We discuss in this section the cost of refining on the cooling length and strategies we envision to mitigate it.

Simulations in our main suite were performed with a different number of cores across different machines. Their total costs thus mix the increased computational load with different intrinsic hardware and parallel efficiencies, making a direct comparison difficult. To provide a cleaner test, we start from our most refined setup at the time shown in Figure~\ref{fig:examplecoolinglength} and~\ref{fig:amr} ($t=540 \, \Myr$; i.e. neither particularly quiescent nor outflowing) and restart four runs for 6 wall-time hours on 768 cores of the same machine, gradually degrading the cooling length refinement target to 36 pc, 72 pc, and none. Table~\ref{table:runs} (right column) reports the median timestep cost (in CPU wall-clock time per simulated Myr) for each of these tests, which we stress can not be extrapolated to estimate the total simulation cost as the timestep strongly varies depending on the outflowing conditions.

As expected, the median cost of a timestep is rising as we increasingly resolve the cooling length, driven by the accordingly increasing number of grid cells (Figure~\ref{fig:amr}). However, this cost only increases by 2 and 28 per cent for our intermediate setups that resolve the cooling length to 72 and 36 pc respectively. It then sharply grows to a 3-fold increase in the most refined run that matches the cooling length with the ISM resolution at 18 pc. 

This non-linear increase reflects the fact that the computational effort is dominated by the densest, most refined cells that have more restrictive Courant–Friedrichs–Lewy conditions and higher convergence costs for the chemistry and radiative transfer solvers. In other words, intermediate setups resolving the diffuse gas twice or four times worse than the ISM are not free, but their computational costs can be kept under control as adding many coarser cells remains a sub-dominant cost compared to computations in the fine ISM gas cells. Given the already large gains in outflow loading factors we observe at 72 and 36 pc (Figure~\ref{fig:loadingfactors}), we foresee these intermediate setups as the most attractive and tractable options for cosmological setups. 

In fact, cosmological simulations of galaxies with such an AMR strategy refining on the cooling length have already been achieved (\citealt{Simons2020, Lochhaas2021VirialTemperature,Lochhaas2022NonHydroEq}), although at a comparatively much lower spatial resolution in both the ISM and the cooling length ($\approx 200 \, \pc$) than here. This lower resolution arises from targeting larger galaxies than the dwarf considered here, and from fundamental differences between the isolated setup of this work and a live cosmological environment. The presence of dense filaments and inflows, of merging galaxies with their stripped tails of gas, and of an already multiphase and turbulent CGM surrounding the central galaxy all add gas highly susceptible to thermal instabilities. This would likely trigger large amounts of additional refinement and sharply increase the computational and memory costs of the simulation. 

Although field-testing is required to pinpoint the achievable cooling-length resolution in a given cosmological setup with \textsc{ramses-rtz}, we remain confident of its overall tractability due to the overwhelmingly dominant cost of the ISM gas. Furthermore, reducing the computational expense is possible by relaxing our requirement that the cooling length is resolved by at least $N_{\text{cells}} = 8$, although further exploration is needed to understand the convergence properties of our simulations with $N_{\text{cells}}$. Another mitigation strategy could also be to release additional resolution levels gradually over cosmic time for the cooling length as is often done for the ISM (e.g. \citealt{Snaith2018}). Or more simply to restart a simulation for a limited time period around a specific event such as a starburst to focus all the effort on one maximally-resolved galactic outflow in a self-consistent environment.  

\subsection{Numerical uncertainties in resolving the multiphase structure of galactic outflows} \label{sec:discussion:convergence}

Our results show that incrementally increasing resolution in the diffuse, outflowing gas boosts both the mass and energy loading factors of outflows as they escape the galaxy. In Section~\ref{sec:outflowstructure}, we tie this to a better-resolved multiphase structure. In this Section, we discuss the physical and numerical mechanisms responsible for this effect, and the remaining uncertainties associated with modelling the hydrodynamics of galactic outflows.

Multiple physical processes are responsible for setting the multiphase structure of gas on small scales, ranging from shocks, hydrodynamical instabilities and cooling and heating processes (Section~\ref{sec:intro} and \citealt{Faucher-Giguere2023} for a review). The primary target of this study is to better capture the poorly-resolved cooling length in the diffuse gas (Figure~\ref{fig:examplecoolinglength}). This can lead to an artificial suppression of multiphase structures (e.g. discussion in \citealt{Hummels2019}). We tackle the issue using the refinement strategy described in Section~\ref{sec:setup:refinement}, but a direct consequence of increasing resolution in the diffuse gas is to also help mitigate other numerical effects. 

In particular, numerical diffusion of hydrodynamical quantities due to the advection of a fluid in motion with respect to a fixed grid (e.g. \citealt{Robertson2010, Teyssier2015}) is particularly important in the context of fast outflowing gas ($\geq 100 \, \kms$). Advection errors are demonstrably reduced with increasing resolution, and our most resolved simulations would thus suffer less from advection-driven numerical mixing between gas phases. Similarly, the heating, lifetime and propagation of shocks are also better captured with increasing resolution (e.g. \citealt{Skillman2008, Creasey2011, Vazza2011}) and typically under-resolved in the diffuse gas (e.g. \citealt{Bennett2020}). A better treatment of shocks then comes as a by-product of refining on the cooling length (e.g. Figure~\ref{fig:fluidvisualization}), improving their ability to thermalize gas at higher temperatures in the lowest density regions which are better sampled. Furthermore, resolving the transition between subsonic and supersonic gas in the outflow is key to accurately capture the acceleration of the hot phase (e.g. \citealt{Chevalier1985, Fielding2022}). This can be under-resolved with traditional schemes (\citealt{Smith2024}), and is improved in our more resolved runs (Figure~\ref{fig:tvz_1kpc}). Lastly, hydrodynamical turbulence is heavily suppressed by insufficient resolution, and sets the characteristic growth timescales of the cold phase (\citealt{Gronke2018, Tan2021, Gronke2022}). The traditional quasi-Lagrangian strategy erases small-scale information in the diffuse gas and likely under-estimates the amount of gas turbulence (see also \citealt{Bennett2020}). The added resolution visibly improves the level of turbulence in the outflowing gas (e.g. in this \href{https://www.youtube.com/watch?v=zQ44XiBkQiA}{movie}), despite our refinement strategy still suppressing turbulence due to refinement/derefinement noise (\citealt{Teyssier2015}, see also \citealt{Martin-Alvarez2022UniGrid} for other strategies to mitigate this issue in a galaxy formation context). 

Pinpointing the respective importance of each of these coupled numerical effects remains difficult in our current setup. Disentangling them could be possible by implementing complementary refinement strategies that target one effect at a time (e.g. shock refinement; \citealt{Bennett2020}) and are used in turn. But we stress that the numerical gains associated with targeting the cooling length of the diffuse gas are multifold, and extend well beyond better-captured thermal instabilities.

Despite these gains however, our simulations do not exhibit signs of numerical convergence in gas temperature, density or velocity PDFs (Figures~\ref{fig:prhos} and~\ref{fig:tvz_5kpc}), highlighting that we are still under-resolving relevant physical processes. This should be expected as resolving their characteristic length scales of each individual microscopic process is required to obtain numerically-converged results and can require sub-pc resolution in the diffuse gas (e.g. \citealt{Koyama2004, Kim2013}). However, the interaction of many processes, each with their own preferred regime and characteristic scales, somewhat blurs this picture, and quantities integrated across the plasma might converge without ever resolving microscopic scales (\citealt{Tan2021}).

In fact, despite the clear growth of outflow loading factors at 5 kpc with increasing resolution (Figure~\ref{fig:loadingfactors}), this growth slows as the resolution keeps improving. This hints that convergence in such integrated quantities could be achieved, in particular in a statistical sense, given the stochasticity associated with star formation activity. More quantitative statements, however, require a wider exploration of numerical resolutions and models.

To summarize, the large boost in outflow loading factors induced by better-resolving the structure of the wind (Figure~\ref{fig:loadingfactors}) strongly motivates extending our analysis. In particular, future studies quantifying the respective roles of the physical and numerical effects discussed above will be key to gain an understanding on the convergence and robustness of our modelling of SN-driven winds, and of their predicted efficiency at regulating galactic star formation. We stress that such studies would need to be complemented by parallel, complementary explorations of the uncertain launching physics of the wind within the ISM. This aspect remains unexplored in this work, but different star formation models, feedback channels and stellar evolution processes can also generate order-of-magnitude variations in dwarf galaxy outflow loading factors (e.g. \citealt{Agertz2020EDGE, Emerick2020StellarFeedback, Smith2020PhotoRT, Andersson2023Inferno, Steinwandel2023Runaways} for recent studies; \citealt{Naab2017} for a review). 

\section{Conclusion} \label{sec:conclusion}
We perform radiative-hydrodynamical numerical simulations of an isolated dwarf galaxy, improving the thermodynamical modelling of the outflowing gas escaping the galaxy. Each simulation drives self-consistent galactic outflows from SNe and photo-ionization feedback within the ISM resolved at $18 \, \pc$, and accounts for the non-equilibrium chemistry and cooling of gas coupled to the radiative transfer using the AMR \textsc{ramses-rtz} code (\citealt{Katz2022RTZ}). 

To remedy the under-resolved cooling length in the diffuse outflow when using a traditional quasi-Lagrangian refinement strategy (Figure~\ref{fig:examplecoolinglength}), we implement a new refinement strategy in \textsc{ramses-rtz} targeting the local gas cooling length computed on-the-fly from the non-equilibrium, line-by-line cooling rate. We then perform additional simulations of our dwarf galaxy incrementally resolving the cooling length down to 72, 36 and 18 pc.

The additional refinement criterion on the cooling length improves the resolution in the diffuse out-of-the-galaxy-plane gas (Figure~\ref{fig:amr}), without significantly modifying the average star formation rate, the amount of stellar mass formed (Figure~\ref{fig:timeevolution}), or the star formation and supernova feedback conditions in the dense, ISM gas (Appendix~\ref{app:similarsfs}). This provides us with a controlled study, where the launching mechanics of the outflows within the galaxy are comparable but their propagation as they escape the galaxy is better resolved.

With increasing resolution, outflows become increasingly multiphase, exhibiting both a systematically higher fraction of colder and denser gas, and a more diffuse phase featuring an increased scatter in temperature (Figure~\ref{fig:trhospdfs} and~\ref{fig:prhos}). Furthermore, our improved numerical scheme better captures the propagation of the hot and energetic phase of the outflow. Preventing artificial mixing between gas phases leads to systematically hotter gas close to the disc, as shock heating become ever more efficient. Furthermore, better sampling the diffuse gas also allows us to resolve the subsonic, hot regions behind shocks and the key transition between supersonic and subsonic gas (Figure~\ref{fig:tvz_1kpc}). These improvements close to the disc then extend to increased energetics at larger radii, with the hot phase staying systematically hotter at large radii (Figure~\ref{fig:tvz_5kpc}). 

Over $\approx 500\, \, \Myr$ of evolution, the combination of these factors leads to a five-fold increase in the average energy loading factor of the galaxy, and two-fold increases in the average mass and metal loading factors in the inner CGM ($r = 5\, \kpc$, Figure~\ref{fig:loadingfactors}). These boosts occur without modifications to the internal feedback budget or modelling and are robust to stochastic realizations of the star formation history (Appendix~\ref{app:stochasticity}). Rather, they stem from better resolving the hydrodynamical and thermodynamical processes as the outflow propagates into the CGM (Section~\ref{sec:discussion:convergence}). The significance of this boost in outflow energetics without new feedback mechanisms strongly motivates future studies extending the analysis to a cosmological context. Quantifying how such outflows interact with a self-consistent, already-multiphase cosmological CGM, rather than the isolated hot halo of this study, will be key to understanding their improved efficiency at regulating galactic star formation over cosmic time.

Furthermore, a key novelty of our simulations is to track the non-equilibrium abundances of $\geq 60$ ions, allowing us to robustly capture the ionic structure of the diffuse gas (Figure~\ref{fig:coveringfractions}). Covering fraction of low-ionization ions (e.g. \hi) exhibit a limited increase, primarily driven by a more prominent dense and colder phase better (e.g. $N_{\text{\hi}} \geq 10^{18}\, \cmsquare$ increasing by 15 per cent). Higher-ionization ions (\ovi, \ovii) show larger enhancements following the hotter temperatures of our outflows with increased resolution, with both $N_{\text{\ovi}} \geq 10^{13}\, \cmsquare$ and $N_{\text{\ovii}} \geq 10^{13}\, \cmsquare$ increasing by $\approx 50$ per cent between our extreme resolutions.

Our results demonstrate the importance of additional resolution in the diffuse gas to capture the ionic structure of outflows, and interpret their observable signatures in emission and absorption. Dedicated companion papers building upon the improved treatment of outflows presented here will quantify the robustness of inferences of outflow properties from observational diagnoses (see also this attached \href{https://www.youtube.com/watch?v=zQ44XiBkQiA}{movie}). In particular, we will post-process our simulations with resonant-line radiative transfer using the \textsc{rascas} code (\citealt{Michel-Dansac2020}) to obtain mock \mgII, \siII$\,$ and \feII$\,$ spectra of our outflows. This will allow us to assess the accuracy with which such observations (e.g. \citealt{Rupke2005Data, Martin2009, Rubin2011, Martin2013}) can recover outflow properties and loading factors (H. Katz et al. in prep). We also plan to use our simulations to assess potential biases in recovering spatially-resolved outflow metallicities from emission line maps (\citealt{Cameron2021}, A. Cameron et al. in prep).

Finally, thermally-unstable and multiphase gas are common features across galaxy formation and astrophysics. The approach presented in this study provides a highly modular AMR strategy to better resolve the cooling length of the diffuse gas, with a computational cost that can be adapted to galaxy formation applications while taking into account the resources at hand (see Section~\ref{sec:discussion:cpucost} for a discussion). We foresee a wide range of potential applications where the combination of this strategy with the non-equilibrium, ion-by-ion cooling of \textsc{ramses-rtz} could provide significant modelling improvements. Namely, to help resolve the multiphase structure of the diffuse, low-density CGM surrounding galaxies (e.g. \citealt{Hummels2019, Peeples2019, Suresh2019, vandeVoort2019, Lochhaas2021VirialTemperature, Lochhaas2022NonHydroEq}), to better capture the potential shattering of cosmological filaments and sheets (\citealt{Mandelker2020,Mandelker2021}), or to provide an efficient way to improve the numerical treatment of cosmological ram-pressure striping (e.g. \citealt{Simons2020}) and better understand the cooling, star-forming tails of jellyfish galaxies (e.g. \citealt{Tonnesen2012, Tonnesen2019, Lee2022}).

\section*{Acknowledgements}
We would like to thank Oscar Agertz and Eric P. Andersson for insightful discussions during the construction of this work and comments on an earlier version of this manuscript. We also thank the referee for a thorough assessment of this work that significantly improved the quality of the manuscript. MR and HK are supported by the Beecroft Fellowship funded by Adrian Beecroft. This work was performed in part using the DiRAC Data Intensive service at Leicester and DiRAC@Durham facilities, operated by the University of Leicester and Institute for Computational Cosmology IT Services, which form part of the STFC DiRAC HPC Facility (www.dirac.ac.uk). These equipments are funded by BIS National E-Infrastructure capital grants ST/K000373/1, ST/P002293/1, ST/R002371/1 and ST/S002502/1, STFC DiRAC Operations grant ST/K0003259/1, and Durham University and STFC operations grant ST/R000832/1. DiRAC is part of the National E-Infrastructure. For the purpose of Open Access, the author has applied a CC BY public copyright licence to any Author Accepted Manuscript version arising from this submission.

We thank the developers of Pynbody (\citealt{Pontzen2013}), Tangos (\citealt{Pontzen2018}), NumPy (\citealt{vanderWalt2011}), SciPy (\citealt{Virtanen2020}), Jupyter (\citealt{Ragan-Kelley2014}) and Matplotlib (\citealt{Hunter2007}) for providing open-source software used in this work. The Astrophysics Data Service (ADS) and arXiv preprint repository were used extensively in this work.

\section*{Author contributions}
The main roles of the authors were, using the CRediT (Contribution Roles Taxonomy) system\footnote{\url{https://authorservices.wiley.com/author-resources/Journal-Authors/open-access/credit.html}}:

MR: Conceptualization ; Data curation; Formal analysis; Investigation; Project Administration; Writing – original draft. HK: Conceptualization; Methodology; Software; Writing – review and editing. AC: Conceptualization; Writing – review and editing. JD: Resources; Writing – review and editing. AS: Resources; Writing – review and editing.

\section*{Data availability}

The data underlying this article will be shared upon reasonable request to the corresponding author.

%%%%%%%%%%%%%%%%%%%%%%%%%%%%%%%%%%%%%%%%%%%%%%%%%%



\bibliographystyle{mnras}
\bibliography{CGM-Cooling-length} 

%%%%%%%%%%%%%%%%%%%%%%%%%%%%%%%%%%%%%%%%%%%%%%%%%%

%%%%%%%%%%%%%%%%% APPENDICES %%%%%%%%%%%%%%%%%%%%%

\appendix
\section{Star formation statistics} \label{app:similarsfs}

We check in this Appendix that the star formation properties of our simulations are unaffected by the additional refinement scheme on the cooling length, ensuring that the stellar distribution and its associated feedback are consistent between runs. 

Figure~\ref{fig:timeevolution} shows that the total amount of stars formed over the course of the simulation is similar to within 15 per cent between runs at different resolutions. We also checked that the stochastic re-simulations at the same resolution presented in Appendix~\ref{app:stochasticity} verify this condition as well. This guarantees that the total, integrated feedback budget over the evolution of our galaxies is largely unchanged, but not that its instantaneous coupling and efficiency is unaffected.

\begin{figure}
  \centering
    \includegraphics[width=\columnwidth]{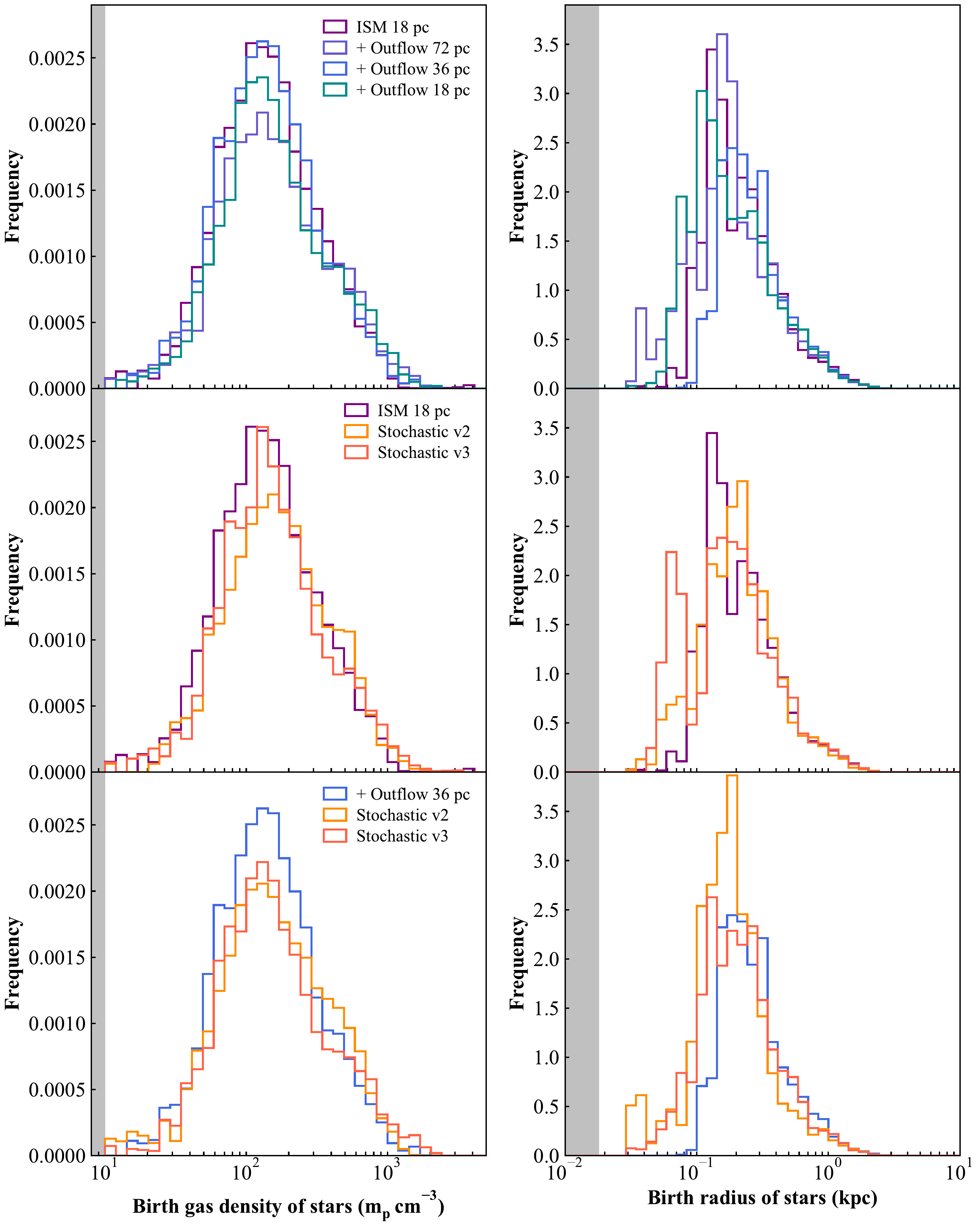}

    \caption{Distribution of gas densities (left) and galactic radii (right) for star particles at their time of birth. Differences between runs with increasingly resolved cooling lengths (top row) are comparable with run to run scatter at a fixed setup (middle and bottom rows), confirming that star formation proceeds similarly in all simulations.}
    \label{fig:starformationproperties}
\end{figure}

A formal quantification of this would require storing the properties of each individual SN to obtain their three-dimensional clustering and explosion conditions (e.g. \citealt{Smith2020PhotoRT}), but this information is unfortunately not recorded by our main simulations. Instead, we save the gas density and galactic spherical radius at which stellar particles are born, providing a useful proxy for the distribution of young stars that are the main feedback contributors. We show their distributions in Figure~\ref{fig:starformationproperties}.

First comparing runs with an increasingly resolved cooling length (top panels), we observe significant differences between their gas density and radii distributions of stellar particles at birth. We compute the two-dimensional Kolmogorov-Smirnov (KS) distance from the reference distribution (`ISM 18 pc') and find (0.06, 0.10, 0.08) for the density distributions and (0.09, 0.10, 0.09) for the galactic radii. Such statistics would be enough to reject that the two samples are drawn from the same underlying distribution at high confidence (p-values $<10^{-7}$). 

This rejection, however, likely arises from the large sample size ($\approx 10,000$) affecting the KS test, rather than from intrinsic differences. In fact, the level of variations between distributions with different cooling lengths is comparable or smaller to run-to-run scatter at a fixed numerical setup (middle and bottom rows). In these cases, we respectively find KS distances of (0.07, 0.09) and (0.04, 0.06) between distribution of gas densities, and (0.06, 0.04) and (0.06, 0.10) between distributions of radii. We also verified that the Wasserstein distances between distributions, which are less sensitive to extrema values, are comparable between stochastic re-simulations and different resolution runs. 

\begin{figure}
  \centering
    \includegraphics[width=\columnwidth]{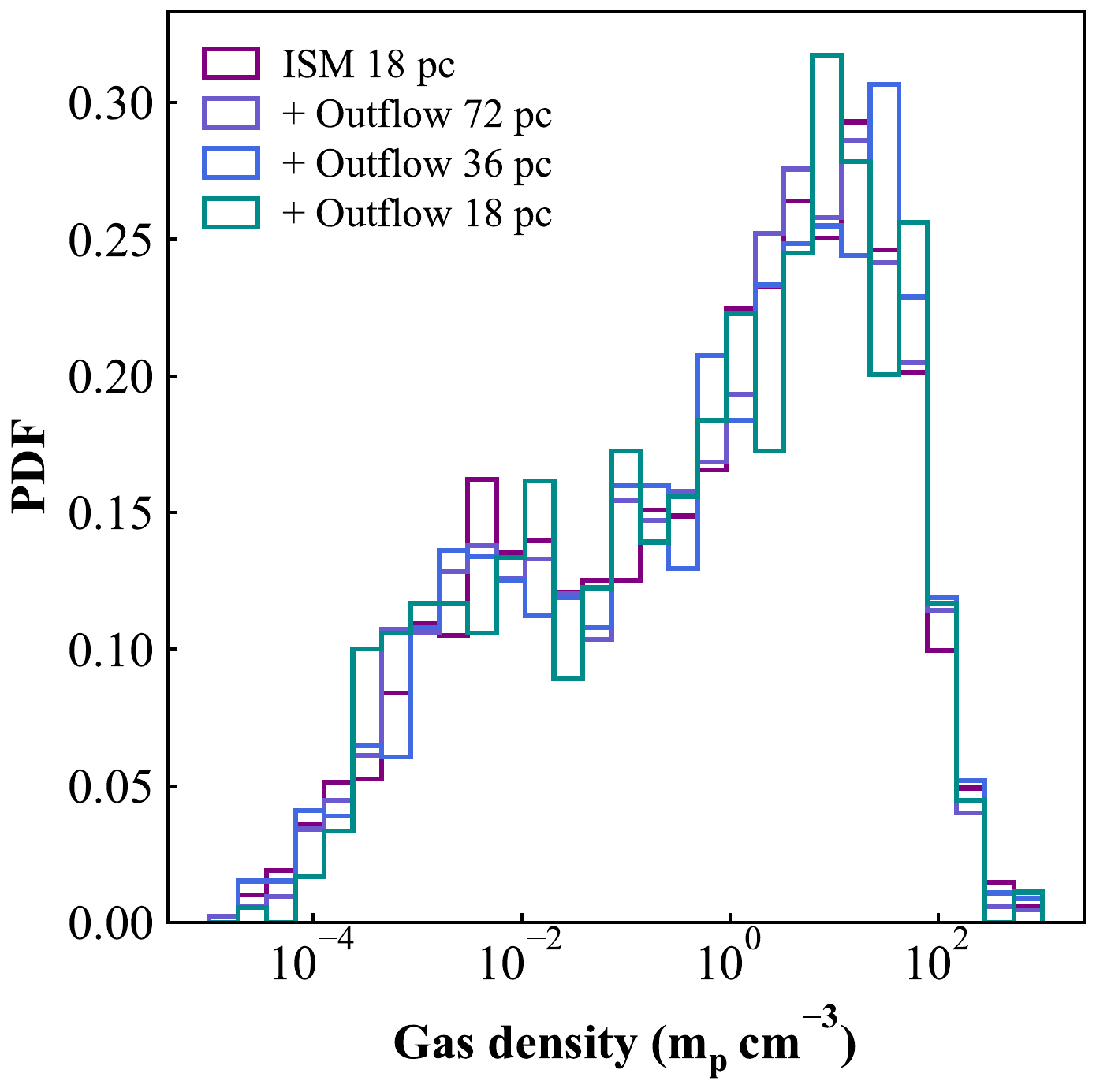}

    \caption{Distribution of gas densities at which SNe explode in the four re-simulation tests described in Section~\ref{sec:discussion:cpucost}. Differences in distributions are minimal, confirming that SN feedback proceeds similarly at all resolutions of the cooling length.}
    \label{fig:snproperties}
\end{figure}

We thus conclude that introducing additional refinement on the cooling length does not impact the star formation within our galaxies further than natural stochasticity, with star formation proceeding at a consistent range of densities and radii between all runs. This could have been expected since we minimally modify the AMR structure of the dense disc where star formation occurs. Even when additional resolution is visible in the galactic plane close to the ISM resolution (Figure~\ref{fig:amr}, right-most), such gas is diffuse and thermally-unstable rather than dense and star-forming, and is being better resolved due to its small cooling length. Our star formation criterion based on convergent flows ensures that star formation proceeds in a similar range of densities and radii in this case (green) than in the other runs. 

To provide an additional verification that feedback conditions in the ISM are comparable between runs, we store the densities within which SNe explode for our four limited-time a-posteriori re-simulations described in Section~\ref{sec:discussion:cpucost}. Figure~\ref{fig:snproperties} shows their distribution across the four resolutions for the cooling length. Differences in these distributions are minimal, giving strong confidence that the star formation, feedback and outflow launching conditions within the ISM are comparable at all resolutions of the cooling length. Rather than differences in initial energetics, the results in Section~\ref{sec:outflowstructure} stem from better resolving the propagation of the outflow into the CGM.

\section{Sensitivity of our results to stochasticity} \label{app:stochasticity}

\begin{figure*}
  \centering
    \includegraphics[width=\textwidth]{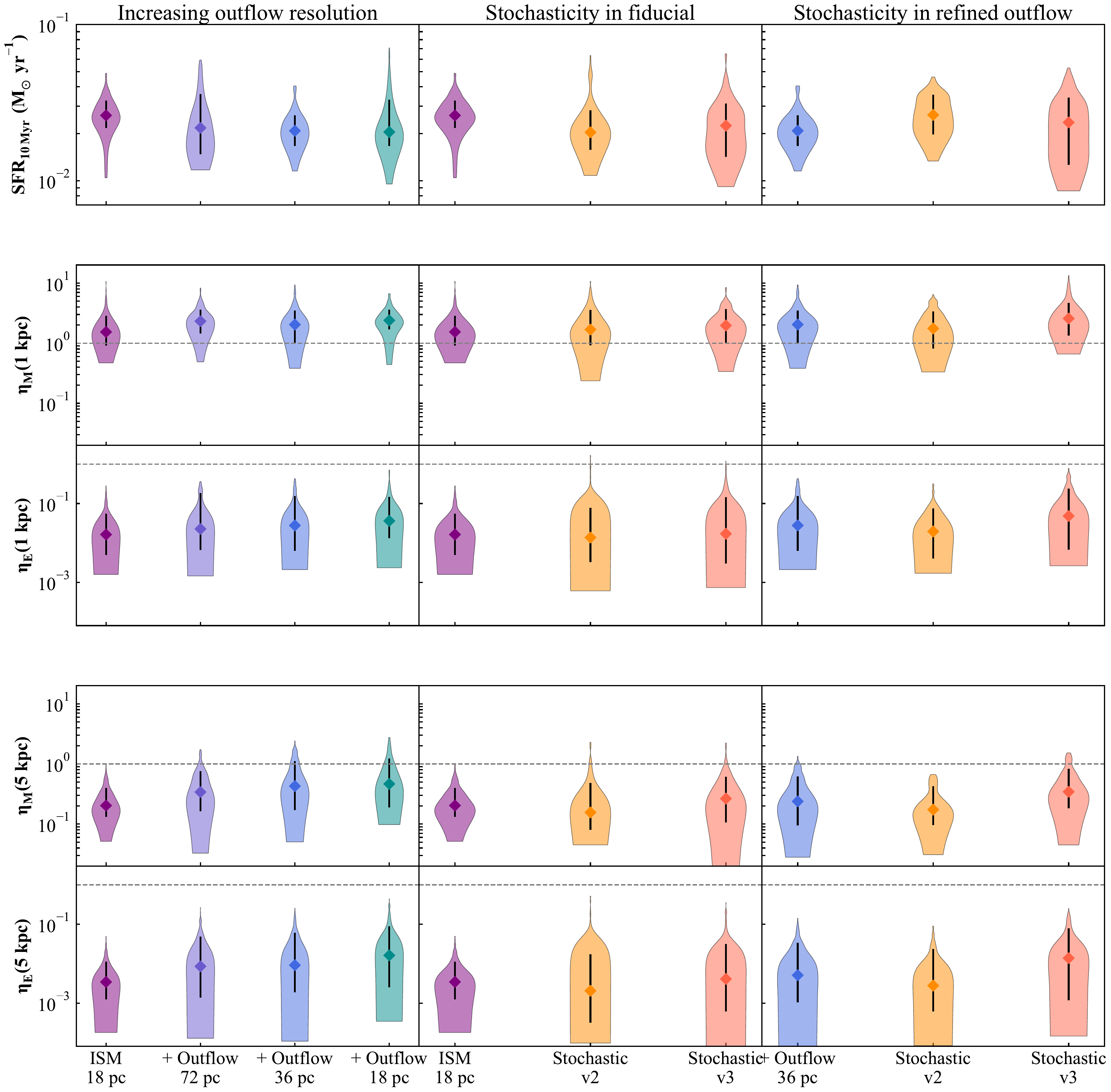}

    \caption{Distribution of star formation rates (top row), mass and energy loading factors at $|z|=1\, \kpc$ (middle rows), and mass and energy loading factors at $r=5\,\kpc$ (bottom rows) for our resolution study (left column), three stochastic re-runs of our fiducial setup (middle column) and three stochastic re-runs refining the cooling length down to 36 pc (right column). All simulations show consistent star formation rates and loading factors close to the disc, confirming that star formation activity and the launching of the winds close to the disc occur similarly in all cases. Larger stochastic shifts are visible further away from the galaxy (bottom rows, middle and right columns), but better resolving the cooling length (right column) always sharply increases mass and energy loading factors compared to the fiducial case (middle column). In fact, stochasticity probed in our fiducial setup is incapable of reproducing the high loading factors obtained at higher resolutions (left and right columns), making it a sub-dominant effect compared to the physical gain of better resolving the multiphase structure of outflows.}
    \label{fig:stochasticity}
\end{figure*}

Although all simulations in the main suite form a similar amount of stars overall (Figure~\ref{fig:timeevolution}), their instantaneous star formation rate and their ability to drive powerful outflows at a given time also depends on more stochastic parameters, such as the instantaneous ISM conditions and clustering of SNe. In this Appendix, we perform additional simulations to understand the level of variance from different star formation realizations, and confirm that results in Section~\ref{sec:outflowstructure} are robust to this stochasticity.

To this end, we first simulate the base fiducial setup (i.e. only with the quasi-Lagrangian refinement scheme) three times, changing only the random seed of star formation to obtain a different evolution at fixed physical setup. We then repeat this exercise with refinement on the cooling length down to 36 pc. Figure~\ref{fig:stochasticity} then compares their distribution of star formation rates (top panels), mass and energy outflow rates at 1 kpc (middle rows), and mass and energy outflow rates at 5 kpc (bottom rows) as defined in Section~\ref{sec:outflowstructure:loadingfactors}. As in Section~\ref{sec:outflowstructure:loadingfactors}, we remove inflows from the presented statistics which only occur for the fiducial setups, making trends when comparing to higher resolution runs more conservative.

Focussing first on each suite of three re-simulations (middle and right columns), we observe visible shifts in the median star formation, mass and energy outflow rates that confirm the presence of stochasticity at fixed numerical setup. Nonetheless, star formation rates have well overlapping 16-84 confidence intervals (black lines) both within each individual suite (top middle or top right), across the two stochastic re-simulation suites (top middle and right), and across the resolution study (top left). This further supports that star formation is regulated similarly across all simulations (see also Appendix~\ref{app:similarsfs}).

Mass and energy outflow rates close to the disc ($|z|=1\, \kpc$; second and third row) are also very stable against stochasticity, with both re-simulations suites (middle panels, middle and right columns) displaying strongly overlapping 16-84 confidence intervals that also overlap with the resolution study (middle, left).

Away from the disc ($r=5\, \kpc$, bottom rows), mass and energy outflow rates show larger differences between stochastic re-simulations (yellow and red), whether in the fiducial case or when refining on the cooling length (middle and right columns, respectively). We also note that higher average star formation rates (e.g. top, yellow) does not necessarily translate to more vigorous outflows on average (bottom, yellow), reflecting the complexity of sustaining galactic winds. Despite these larger shifts, the stochasticity probed in our fiducial setup (middle column) is incapable of reproducing the sharp increase in outflow loading factors at 5 kpc observed when increasing numerical resolution out-of-the-plane (left and right columns). Even our most vigorous fiducial case (middle column, red) has mass and energy loading factors a factor $\approx 3$ short of the quietest run at 36 pc (right column, yellow) and is even further away from other resolved runs (left).

We thus conclude that stochasticity at a fixed numerical setup impacts quantitative determinations of outflow loading factors, but that in our case, stochastic differences are sub-dominant to the physical gain of better resolving the multiphase treatment of the outflow. In particular, even fairly limited improvements (e.g. violet in left column) bring a significant gain in loading factors that already outweigh stochasticity without drastically impacting the computational cost (Section~\ref{sec:discussion:convergence}).

\section{Outflow opening angle as a function of resolution} \label{app:openingangle}

Figure~\ref{fig:fluidvisualization} shows a greater opening of the velocity streamlines with resolution at a single time. Here, we quantify how outflow opening angles vary as a function of time and respond to increasing resolution.

\begin{figure}
  \centering
    \includegraphics[width=\columnwidth]{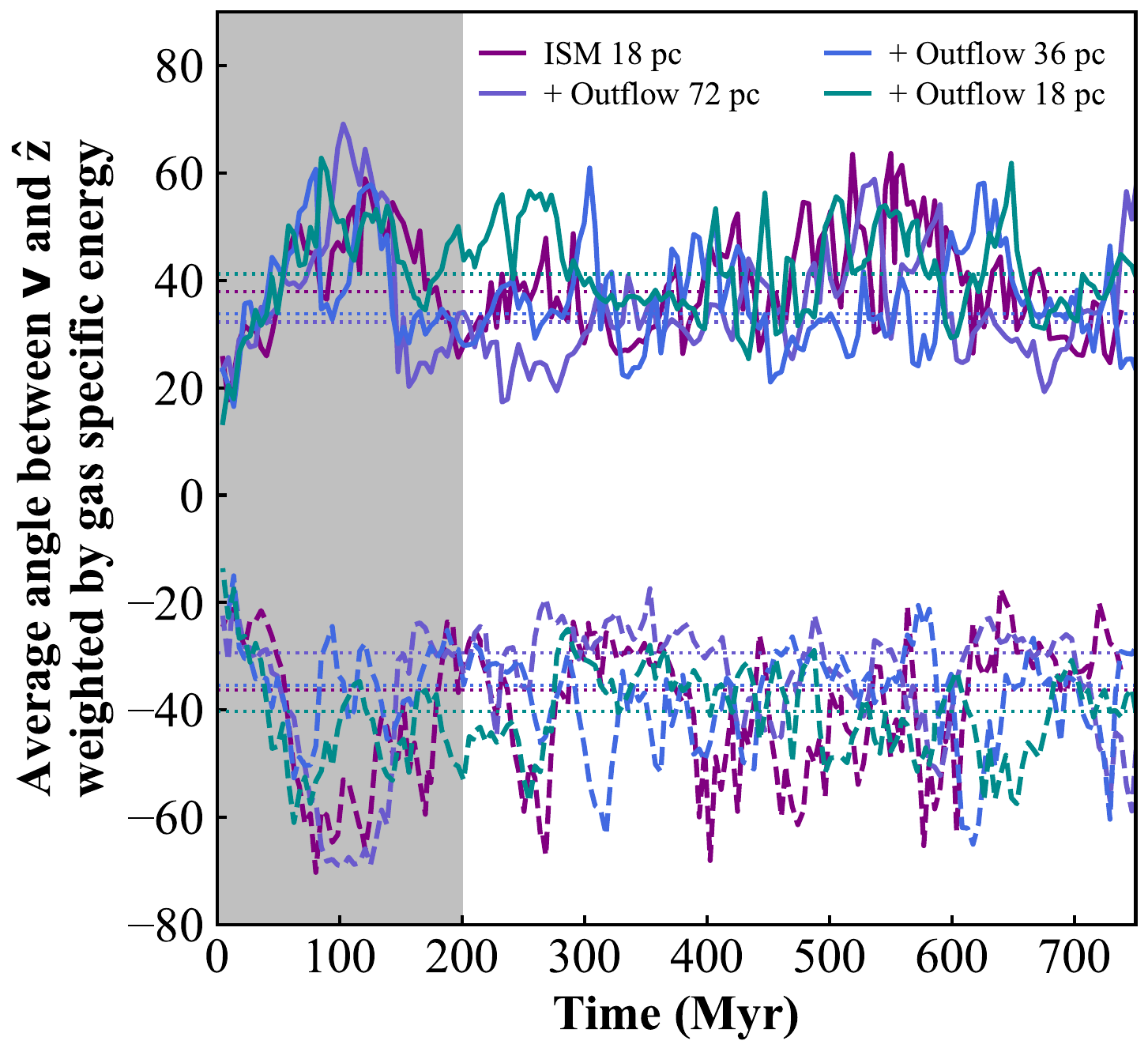}

    \caption{Time evolution of the opening angle of the hot outflowing gas. We show the average angle of the velocity with the z-direction weighted by specific energy, for the same gas as in Figure~\ref{fig:prhos}. Outflows at the highest resolution (cyan) are systematically more open (dotted line showing medians, see also Figure~\ref{fig:fluidvisualization}), in both the upper and lower outflows (solid and dashed, respectively). But there is considerable scatter around these medians following the varying outflow conditions (see also Figure~\ref{fig:timeevolution}), making a quantitative and conclusive link difficult to establish.
    }
    \label{fig:openingangle}
\end{figure}

To illustrate this, we select gas in upper and lower cylinder above and below the galaxy as in Section~\ref{sec:outflowstructure:thermodynamics} ($R\leq 4.0 \, \kpc$ and $0.5 \leq |z| \leq 5 \, \kpc$). For each gas cell, we then compute the angle of their velocity vector and the z-direction. Figure~\ref{fig:openingangle} then shows the time series of the average angle for this gas, weighted by specific energy (to upweight the hot and fast gas). Solid and dashed lines show the upper and lower outflows, and dotted lines show the median at each resolution. 

We find that, at the highest resolution (cyan), the velocity structure is systematically more open (median towards larger angles) in line with the qualitative findings of Figure~\ref{fig:fluidvisualization} and aligning with the increased outflow energetics (Section~\ref{sec:outflowstructure:kinematics}). However, there is considerable scatter in opening angles as a function of time, with all different resolutions producing more or less open velocity structures with varying outflow conditions. In fact, the median opening angle does not respond linearly with improving cooling-length resolution (`Fiducial' in purple has the second-most open outflow, but worse off-the-plane resolution). Furthermore, differences between upper (solid) and lower (dashed) outflows are comparable to the shifts with resolution. A causal link between our improved numerical treatment and systematically more open velocity streamlines thus remains tentative, and we hypothesize that repeating this analysis with a longer time baseline, or in systems with higher mass outflow rates might help establish this result.

\section{Different outflow geometries at 5 kpc} \label{app:fullshell}

\begin{figure*}
  \centering
    \includegraphics[width=\textwidth]{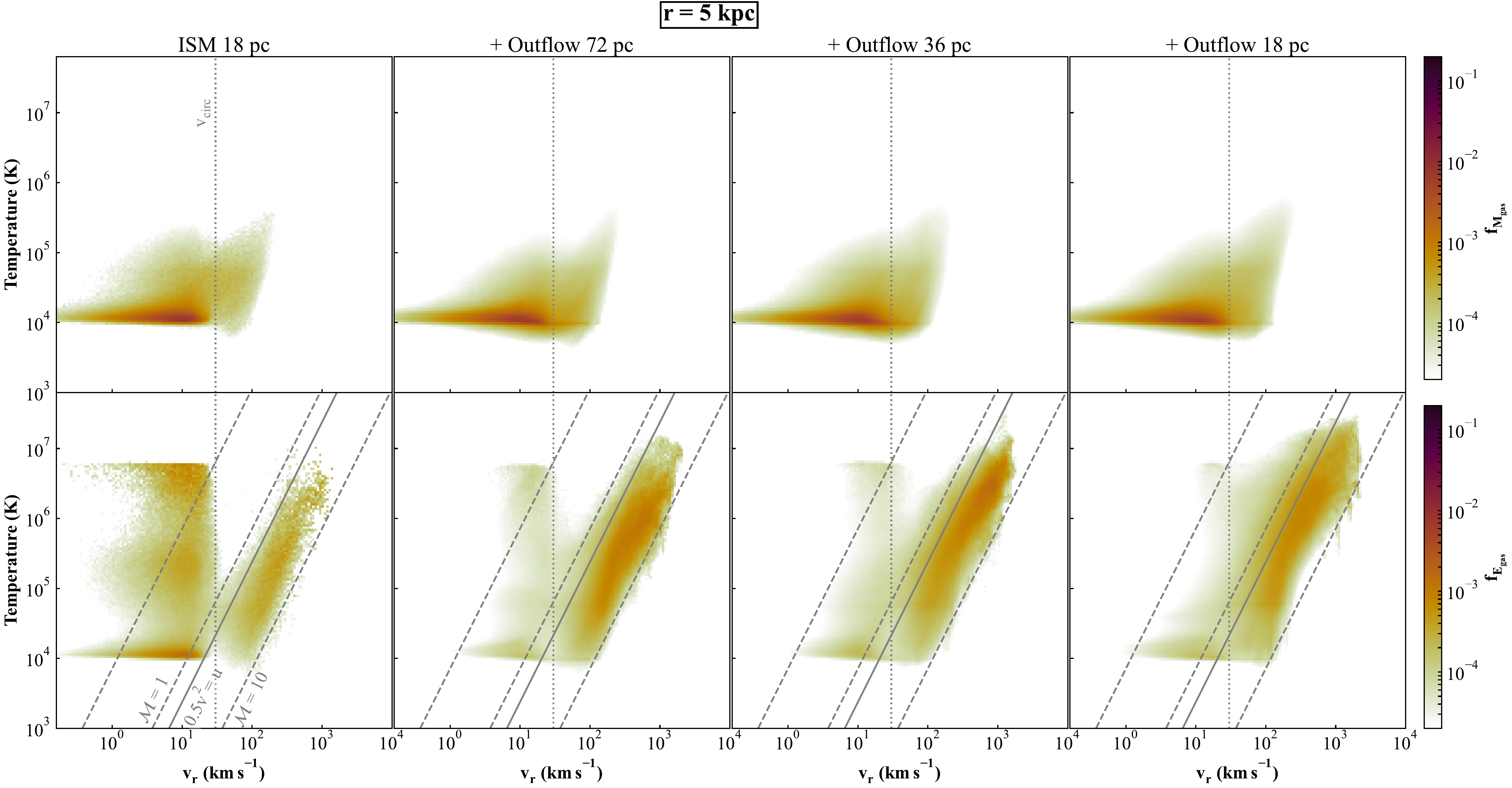}

    \caption{Same as Figure~\ref{fig:tvz_5kpc} but for gas within the full 2-kpc wide shell at $r=5\, \kpc$ rather than restricting to opening angles $\theta=45^{\circ}$. We recover similar features and trends with resolution as in Figure~\ref{fig:tvz_5kpc}, with the addition of a cloud of hot and slow-moving gas visible at all resolutions but particularly prominent in our lower resolution case (bottom, left). This feature arises from the hot hydrostatic halo of the initial conditions of our galaxy, which we aim to remove in our fiducial analysis.
    }
    \label{fig:tvz_5kpc_full_shell}
\end{figure*}

Here, we test the importance of our chosen biconical shell geometry at 5 kpc. We first show in Figure~\ref{fig:tvz_5kpc_full_shell} the same as in Figure~\ref{fig:tvz_5kpc}, but using the full radial shell rather than restricting on polar angles as for the fiducial case.

\begin{figure*}
  \centering
    \includegraphics[width=\textwidth]{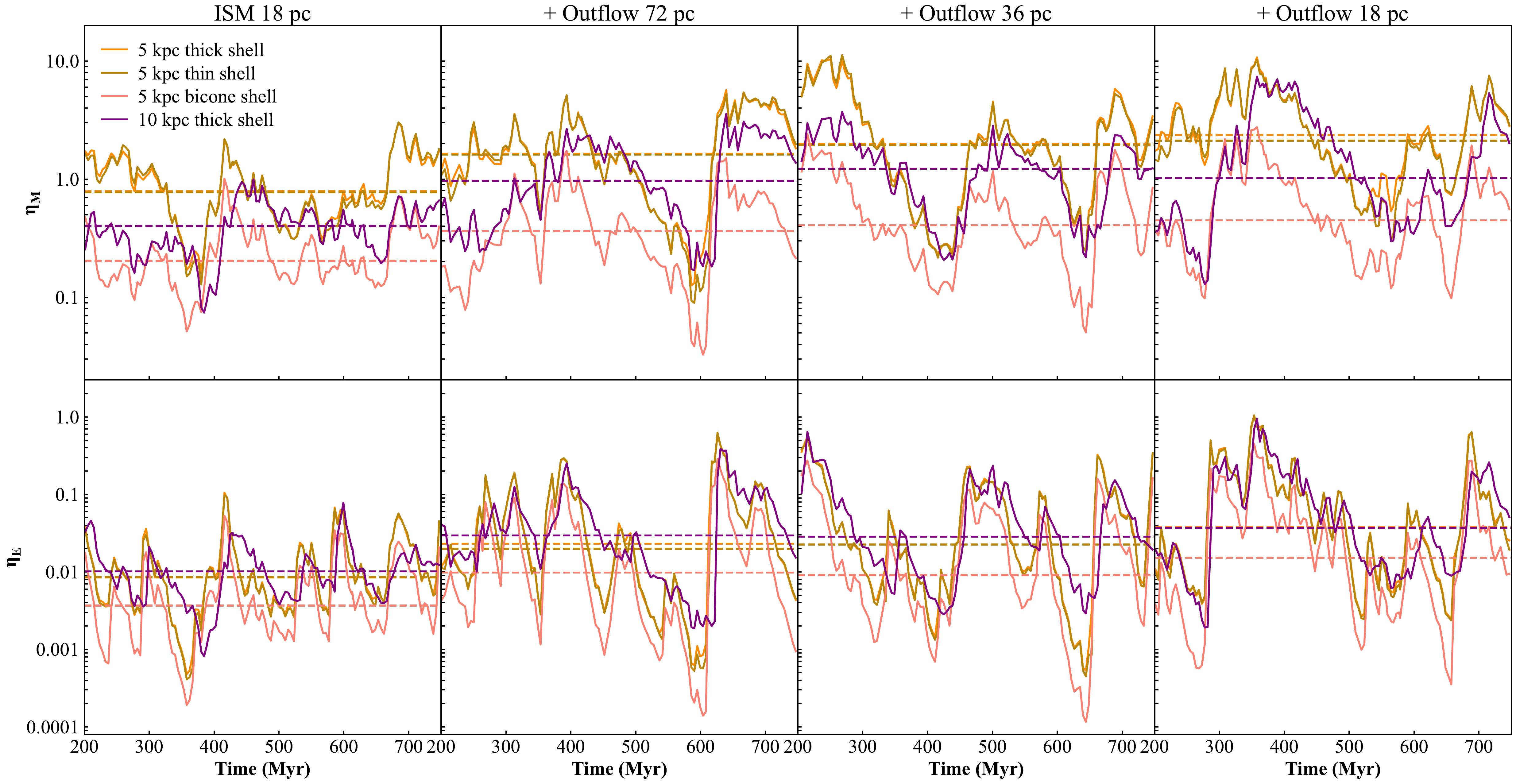}

    \caption{Time evolution of the mass (top) and energy (bottom) outflow loading factors measured in different ways. Whether using a thick (orange), thin (gold), biconical (red) shell placed at 5 kpc or 10 kpc (purple), the outflow loading factors climb with improving resolution in the diffuse gas (left to right).
    }
    \label{fig:5kpc_multiple_outflow_rates}
\end{figure*}

Figure~\ref{fig:tvz_5kpc_full_shell} recovers all the trends observed in Figure~\ref{fig:tvz_5kpc}, in particular the energetic hot outflowing phase being hotter and faster moving at increasing resolution. But using the full radial shell highlights another feature: slow moving gas spreading between two defined tracks at $T = 10^4\, K$ and $T\approx\xScientific{5}{6}\, K$. Rather than the outflow originating from the galaxy, this gas is part of the hot hydrostatic halo of our galaxy, as evidenced by its sharp drop-off at $v = \vcirc$ (we also verified that this gas defines a linear track in the T-$\rho$ plane). Our choice of hot halo conditions are fully dictated by the choice of initial conditions for our isolated galaxy set-up, and its contribution can be significant when the diffuse gas is poorly resolved (bottom, left). In our fiducial analysis, we thus aim to mitigate its importance by selecting a biconical shell to more cleanly separate this gas from SNe-powered outflow which we aim to study.

In Figure~\ref{fig:5kpc_multiple_outflow_rates}, we verify that trends of outflow loading factors with resolution are unaffected by this choice. We show the time series of the outflow mass and energy loading factors (top and bottom) at each of our resolutions (left to right) computed through our fiducial 2-kpc thick shell at 5 kpc with and without its biconical definition (orange and red), a thinner 0.8-kpc shell (gold), and a 2-kpc thick shell at 10 kpc (brown).

In all cases, the outflow loading factors climb with improved resolution (left to right), validating that our trends are robust to these definitions. As expected due to geometrical factors, the mass loading factors defined through the biconical shell are always lower than the full shell, and mass loading factors decrease between 5 and 10 kpc (top, orange versus purple). However, perhaps surprisingly, the energy loading factors increase with radius (bottom) despite the lack of off-the-plane energy injection from stars (e.g. from runaway stars). This arises due to the rising contribution from the hot $T\approx10^6\, \K$ hydrostatic halo, that acts as heat bath in which the galaxy is embedded. This hot halo can still be pressurized by the SNe-powered outflows and slowly expand, but most of its properties are dictated by our choice of initial conditions. To avoid this issue, we thus select a biconical geometry for our fiducial analysis in this study, and plan to extend our work to a cosmological context studying better-resolved inflows and outflows in a self-consistent environment.

%%%%%%%%%%%%%%%%%%%%%%%%%%%%%%%%%%%%%%%%%%%%%%%%%

% Don't change these lines
\bsp	% typesetting comment
\label{lastpage}
\end{document}